\documentclass[aps,pra,twocolumn,groupedaddress,amsmath,amssymb,nofootinbib]{revtex4-1}

\usepackage[colorlinks,pdfusetitle,urlcolor=blue,citecolor=blue,linkcolor=blue,bookmarksnumbered,plainpages=false]{hyperref}

\usepackage{color}

\usepackage{bm}
\usepackage{nicefrac}
\usepackage{siunitx}

\usepackage{graphicx}
\usepackage{bm}
\usepackage{braket}
\newcommand{\pd}[2]{\frac{\partial#1}{\partial #2}}
\newcommand{\plus}[1]{#1_{+}}
\newcommand{\minus}[1]{#1_{-}}

\newcommand{\bx}{\hat{b}_x^{\vphantom{\dagger}}}
\newcommand{\bxd}{\hat{b}_x^{{\dagger}}}
\newcommand{\by}{\hat{b}_y^{\vphantom{\dagger}}}
\newcommand{\byd}{\hat{b}_y^{{\dagger}}}
\newcommand{\bL}{\hat{b}_L^{\vphantom{\dagger}}}
\newcommand{\bLd}{\hat{b}_L^{{\dagger}}}
\newcommand{\bR}{\hat{b}_R^{\vphantom{\dagger}}}
\newcommand{\bRd}{\hat{b}_R^{{\dagger}}}

\newcommand{\B}[1]{\mathbf{#1}}

\begin{document}

\title{Anomalous magnetic moment of an electron near a dispersive surface}

\author{Robert Bennett and Claudia Eberlein}

\affiliation{Department of Physics \& Astronomy, University of Sussex, Falmer, Brighton BN1 9QH, UK}

\date{\today}

\begin{abstract}
Changes in the magnetic moment of an electron near a dielectric or conducting surface due to boundary-dependent radiative corrections are investigated. The electromagnetic field is quantized by normal mode expansion for a nondispersive dielectric and an undamped plasma, but the electron is described by the Dirac equation without matter-field quantization. Perturbation theory in the Dirac equation leads to a general formula for the magnetic moment shift in terms of integrals over products of electromagnetic mode functions. In each of the models investigated contour integration techniques over a complex wave vector can be used to derive a general formula featuring just integrals over transverse electric and transverse magnetic reflection coefficients of the surface. Analysis of the magnetic moment shift for several classes of materials yields markedly different results from the previously considered simplistic `perfect reflector' model, due to the inclusion of physically important features of the electromagnetic response of the surface such as evanescent field modes and dispersion in the material. For a general dispersive dielectric surface, the magnetic moment shift of a nearby electron can exceed the previous prediction of the perfect-reflector model by several orders of magnitude.  \end{abstract}

\pacs{}

\maketitle
\section{Introduction}
The coupling of the quantized electromagnetic and electron fields to each other gives rise to radiative corrections in quantum electrodynamics. One of the quantities altered by these radiative corrections is the electron's magnetic moment, which is particularly interesting because it can be measured to staggeringly high precision \cite{Odom, Hanneke}. The presence of material boundaries affects the fluctuations of the electromagnetic field, thus alters radiative corrections, and thereby causes the magnetic moment for an electron near a surface to differ from its value in free space \cite{Fischbach, Svozil, Bordag, BoulwareBrown, KreuzerSvozil, Kreuzer, BartonFawcett}. This could in principle be an important effect to consider since precision $g$ factor measurements of leptons not only provide stringent tests of quantum electrodynamics but also potentially open up a low-energy route to testing physics beyond the standard model. 

Previous literature on this subject has shown that measurement of the shift is not within the reach of contemporary experiments. However, all these previous investigations have made the crude simplification that the surface may be regarded as perfectly reflecting.  This simplifies calculations, but the perfect-reflector model has obvious physical deficiencies: it does not account for electromagnetic field modes that are evanescent outside the medium, neither does it reproduce the fact that any real medium becomes transparent at high frequencies. Thus, to make a realistic prediction of the potential measurability of this effect, one needs to consider a surface which is imperfectly reflecting and dispersive.

The idea that the use of a more realistic model of the material may significantly affect the shift is not suggested by the results of other electrodynamic boundary-dependent effects. For example, in the calculation of the Casmir-Polder force on an atom in front of a surface \cite{CP}, the perfect reflector model is completely adequate for estimates and its results are reproduced in the expected limiting cases, e.g. taking the refractive index of a nondispersive medium to infinity \cite{Wu} or the plasma frequency of a plasma surface to infinity \cite{BabikerBarton}. The situation is strikingly different for the spin magnetic moment of an electron near a surface where, as we will show, different models for the electromagnetic response of the surface give drastically different results, not necessarily obtainable as limiting cases of each other. 
Something similar has already been observed for the self-energy shift of a free electron near a nondispersive surface \cite{EberleinRobaschik} and has since been investigated also for dispersive dielectric and conducting surfaces \cite{massshift}. The reason for these disagreements between different models has been found to be related to the fact that a free particle admits excitations of arbitrarily low frequency which are dealt with very differently in the various models. However, the low-frequency part of the photon spectrum turns out to dominate the shift, so that one necessarily gets completely different results for the self-energy in models with different low-frequency electromagnetic response. By contrast, for bound atomic electrons the gap to the nearest energy level provides a natural low-frequency cutoff for electronic excitations, so that for the Casimir-Polder shift due to a surface any difference in the low-frequency behaviour of the model of the electromagnetic response of the surface turns out not to matter to leading order.

\begin{figure}
\includegraphics[width = 8cm]{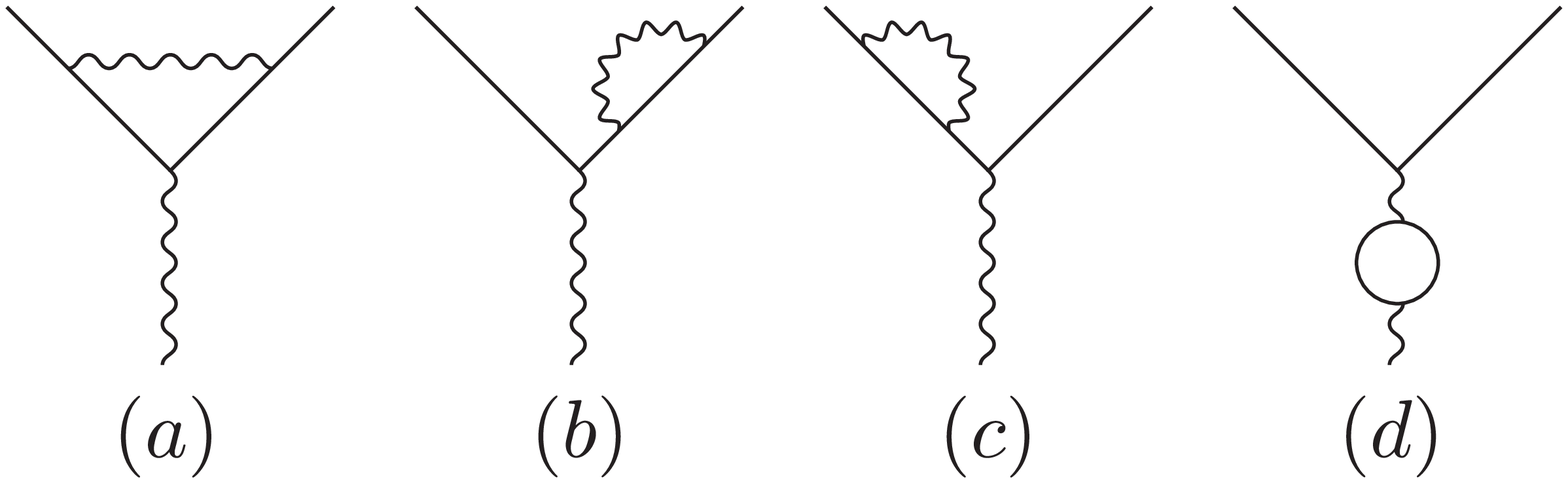}
\caption{\label{fig:FeynmanDiagrams} One-loop reducible and irreducible corrections to the bare vertex}
\end{figure}

\begin{figure}[t]
\includegraphics[width = 8cm]{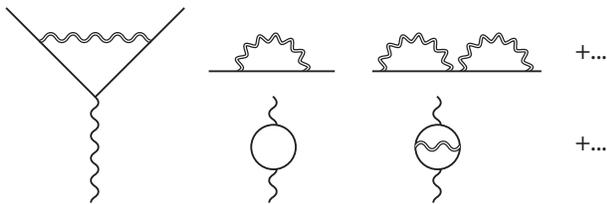}
\caption{\label{fig:ExtraFeynmanDiagrams} Possible $n$ loop insertions to the external legs of the vertex diagram. A double line represents a boundary-dependent propagator}
\end{figure}

In order to determine the quantum-field theoretical corrections to the electron's magnetic moment, one usually calculates the vertex diagram to the respective order of interest [cf. Fig.~\ref{fig:FeynmanDiagrams}(a) to one-loop order]. The reducible one-loop diagrams [cf. Figs.~\ref{fig:FeynmanDiagrams}(b)--(d)] are taken care of through mass and charge renormalization. 
Thus, in free space the evaluation of the magnetic moment to one-loop order $e^2\equiv\alpha$ is a straightforward calculation and graduate textbook material (c.f. e.g. \cite{Peskin}).
However, this changes radically if the electromagnetic field is subject to reflection by a surface. Then even one-loop calculations of quantum electrodynamics get rather complicated \cite{PRDEberleinRobaschik}, firstly because of the loss of translation invariance, and secondly because of the localization of the electron that is required if the calculations are to make physical sense and be interpretable. The photon propagator becomes boundary dependent, so that standard mass and charge renormalization are no longer applicable. Mass renormalization is affected at all orders, and charge renormalization from the two-loop level up, as illustrated by Fig.~\ref{fig:ExtraFeynmanDiagrams}. 

Most of such technical problems can be avoided by taking a different, more appropriate approach. As we seek only the \emph{correction} to the magnetic moment due to the presence of the surface, it follows that any boundary-independent contributions can be discarded in our calculation. The bare electron propagator is not affected by the presence of boundary, provided the electron is at least a few Compton wavelengths away from the surface of the medium so that the interaction between them is wholly electromagnetic, which is the case we are interested in here. Then the one-loop vacuum polarization diagram can be discarded when coupled to a static external field, and to one-loop order the charge and its renormalization are the same as in free space. None of the three remaining one-loop diagrams contain any particle-antiparticle loops.
 Consequently, for the calculation of the \emph{boundary-dependent} corrections to the magnetic moment at one-loop level, it suffices to work with a \emph{first}-quantized electron interacting with a second-quantized photon field. This allows us to borrow well-tried techniques from quantum optics. We shall use perturbation theory to determine the energy shift $-\boldsymbol{\mu}\cdot{\bf B}_0$, due to the presence of the material surface, of the electron's spin in a weak external magnetic field ${\bf B}_0$, and then extract the magnetic moment as the coefficient $\boldsymbol{\mu}$ of the term linear in ${\bf B}_0$. 

In the next section, we give the derivation of the shift in terms of mode functions of the electromagnetic field. Then, Sec.~\ref{nondispmodes} deals with the electromagnetic mode functions for a nondispersive dielectric, Sec.~\ref{PlasmaModes} with those for a plasma model, and Sec.~\ref{DispDiel} with the case of a dispersive dielectric. 
In Sec.~\ref{Integral}, we show in detail how to evaluate the required integrals for the various different models, and in Sec.~\ref{Results}, we discuss the results of the various models. Finally, in Sec.~\ref{ExpRel} we discuss the experimental relevance of our results. We use natural units with $c=1=\hbar$ and $\epsilon_0 = 1 = \mu_0$ throughout. 

\section{Perturbation Theory} \label{PerturbationTheory}

Our starting point is the Dirac equation coupled to an electromagnetic field $A_\mu$
 \begin{equation}
[-i\gamma^\mu (\partial_\mu + ie A_\mu) + m] \psi = 0.
\end{equation}
In order to best exploit the similarities with quantum optical problems we shall use the Dirac equation in its non-covariant form 
\begin{equation}
i\frac{\partial }{\partial t} \psi = [\bm{\alpha} \cdot (\B{p}-e\B{A}) +e \Phi + \beta m] \psi
\label{Dirac}
\end{equation}
where $\gamma^0 = \beta$, $\gamma^i = \beta \alpha^i$, and $A_\mu=( \Phi, -\B{A})$. In order to derive the energy shift of a stationary electron in a weak and static magnetic field and close to a reflecting surface, we aim at a non-relativistic approximation. A routine method of achieving this is to approximate the Dirac equation by means of a Foldy-Wouthuysen transformation, as has been done for working out the magnetic-moment correction of the electron near a perfect reflector \cite{BartonFawcett}. However, this approach requires great care in all its steps since several successive orders in the non-relativistic expansion turn out to contribute to the shift. Additionally, in the only comparable previous literature \cite{BartonFawcett} a second unitary transformation is applied to the transformed Hamiltonian (an approach that the authors term the `Paris Method'). This second transformation only works for the perfect reflector, meaning that a straightforward extension of Ref.~\cite{BartonFawcett}'s work to other surfaces is not possible. Hence we adopt a more generally applicable approach. We work directly with the Dirac equation and apply perturbation theory using the solutions of the Dirac equation in a static magnetic field introduced in Ref.~\cite{JohnsonLippmann} as unperturbed 4-spinor states.

\begin{figure}
\includegraphics[width = 8cm]{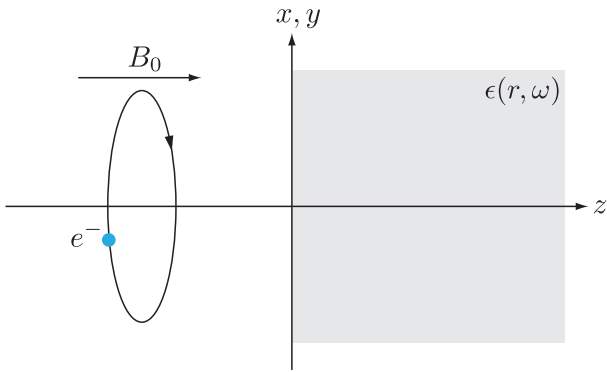}
\caption{\label{fig:HalfSpace}(Color online) Basic setup}
\end{figure} 
 
We subject the electron to a constant classical magnetic field $\mathbf{B}_0$ acting initially the along $\hat{z}$ axis; $\mathbf{B}_0 = B_0 \hat{z}$, as shown in Fig.~\ref{fig:HalfSpace}. Generalization of our results to include the case of magnetic fields parallel to the surface will be provided later. A suitable classical vector potential that generates such a field $\B{B}_0$ is $\mathbf{A_0}=-\frac{1}{2} (\mathbf{r} \times \mathbf{B}_0 )$. Added to this is the quantized photon field $\mathbf{A}_Q$, so that the total vector field entering the Dirac equation (\ref{Dirac}) is
\begin{equation}
\mathbf{A} = \mathbf{A}_0 + \mathbf{A}_Q\,. 
\end{equation} 
The Dirac equation for a particle in a constant magnetic field can be solved exactly, whence the unperturbed Hamiltonian is
\begin{equation}
H_0  =\bm{\alpha} \cdot \bm{\pi} +\beta m\qquad  \mbox{with}\quad \bm{\pi}=\B{p}-e\B{A}_0\;.
\label{unpert_H}
\end{equation}
We show in Appendix \ref{LandauQ} how the eigenstates of the Schr\"{o}dinger Hamiltonian for a particle in a constant magnetic field can be obtained and how one can easily generate those of the corresponding Dirac Hamiltonian from them. The Hamiltonian can be expressed in terms of the creation and annihilation operators for right-circular quanta that serve to move states from one Landau level to the next and obey the bosonic commutation relations
\begin{equation}
[\hat{b}_R, \hat{b}_R^\dagger]=1\ , \qquad [\hat{b}_R,\hat{b}_R]=0=[\hat{b}_R^\dagger, \hat{b}_R^\dagger]\;.
\end{equation}

The perturbation (Dirac) Hamiltonian that describes the interaction of the electron with the quantized field is
\begin{equation}
H_{\text{int}}=-e \gamma_0 \bm{\gamma} \cdot \mathbf{A}_Q=-e\bm{\alpha} \cdot \B{A}_Q\;.
\end{equation}
We note that the scalar component $\Phi$ of the quantized field shifts all states uniformly, so has no impact on the magnetic moment. The energy shift is given by second order perturbation theory as:
\begin{equation}
\Delta E =e^2 \sum_{\Psi_e'} \int d^3\B{k} \frac{|\bra{\Psi_e',1_{\B{k,\lambda}}}\bm{\alpha}\cdot\mathbf{A_Q}\ket{\Psi_e,0}|^2}{E-E'}
\label{E-shift}
\end{equation}
where $1_{\B{k,\lambda}}$ indicates a one-photon state with wave vector $\B{k}$ and polarization $\lambda$, and $\Psi_e$ represents the state of the electron coupled to the classical field $A_0$. The Dirac eigenstates $\Psi_e$ for each Landau level may be written in terms of the corresponding Schr\"{o}dinger eigenstates $|\nu\rangle$, as shown in Appendix \ref{LandauQ}.

The quantized electromagnetic field is written in terms of mode functions $\mathbf{f}_{\mathbf{k}\lambda}$
\begin{equation}
\mathbf{A}_Q = \sum_\lambda \int d^3\B{k}\; (\mathbf{f}_{\mathbf{k}\lambda} \hat{a}_{\mathbf{k} \lambda}e^{-i\omega t}
+\mathbf{f}_{\mathbf{k}\lambda}^* \hat{a}_{\mathbf{k} \lambda}^\dagger e^{i\omega t}) 
\label{A_expansion}
\end{equation}
where $\hat{a}_{\mathbf{k} \lambda}$ and $\hat{a}_{\mathbf{k} \lambda}^\dagger$ are the photon annihilation and creation operators for each wave vector and polarization. Any normalization constants that may appear for the modes in different classes of models have been absorbed into the functions $\mathbf{f}_{\mathbf{k}\lambda}$ in order to preserve the generality of the expressions to be derived. 

Substituting the quantized vector field $\mathbf{A}_Q$ from Eq.~(\ref{A_expansion}) and the Dirac eigenstate $\Psi_e$ as derived in Appendix \ref{LandauQ} into Eq.~(\ref{E-shift}), we have for the energy shift of an electron in Landau level $\nu$ and spin state $s$:
\begin{widetext}
\begin{equation}
\Delta E = e^2 \sum_{\nu ',s'} \sum_\lambda \int d^3\B{k} \frac{|\bra{\nu ,s}(H_0+E_\nu )\boldsymbol{\alpha}\cdot \mathbf{f}_{\mathbf{k}\lambda}(H_0+E_{\nu '})\ket{\nu ',s'}|^2}{4(E_\nu -E_{\nu '}-\omega)E_\nu (E_\nu +m)|E_{\nu '}|(|E_{\nu '}|+m)}  \label{PertMainEqn}
\end{equation}
\end{widetext}
where $\omega$ is the photon frequency. As we are aiming for the energy level shift of a localized stationary or slow-moving electron, we perform a non-relativistic expansion which implies that the electron's rest mass is much bigger than all other relevant energy scales. In this sense we expand, loosely speaking, in powers of $1/m$, and we shall keep contributions to the magnetic moment up to order $1/m^3$. For the surface-dependent corrections to the magnetic moment that we are interested in, this is in fact an expansion in $1/(mz)$, which is the ratio of the electron's Compton wavelength to the distance $z$ of the electron from the surface, and thus very small indeed for all even remotely realizable values of the distance $z$.
The above expression contains terms that are of order $1/m^2$ and higher; however, it turns out that none of the $1/m^2$ terms contribute to the magnetic moment, all the contributions to which are of order $1/m^3$ and higher. We note that fourth-order perturbation theory can contribute only terms of order $1/m^4$ or higher, and therefore does not need to be considered here. 

Since the spin magnetic moment is obtained from the coefficient of the terms of $\Delta E $ that are linear in $\sigma_z B_0$, one needs to carefully enumerate all the possible effects which may generate a dependence on $B_0$. With this in mind, we find that we must go beyond the dipole approximation for the field $\mathbf{A_Q}$, because the multipole expansion
\begin{equation}
\mathbf{A}_Q(\mathbf{r}) = \mathbf{A}_Q(\mathbf{r_0}) + [(\mathbf{r}-\mathbf{r}_0)\cdot\nabla]\mathbf{A}_Q(\mathbf{r_0})+\ldots
\label{multipole}
\end{equation}
contains powers of the displacement operator $\mathbf{r}-\mathbf{r}_0$, whose matrix elements are proportional to $B_0^{-1/2}$ [see Eq.~(\ref{position_elements}) in Appendix \ref{LandauQ}]. For almost all contributions to the energy shift we shall find that the dipole approximation works and we need to take along only $\mathbf{A}_Q(\mathbf{r_0})$, but there is going to be just one exception for which we shall have to take along the next term in the above expansion. This may seem unusual from a view point of quantum optics, but this is ultimately due to the entirely classical effect of the magnetic field causing particle trajectories to bend. We shall see that to order $1/m^3$ there is only one term for which the curvature of the trajectory matters and the variation of the quantum field $\mathbf{A}_Q$ along the curved trajectory causes transitions between one Landau level and the next.

Despite us being interested in the energy shifts only for particles, not anti-particles, we need to sum in Eqs.~(\ref{E-shift}) and (\ref{PertMainEqn}) over all intermediate states, both particle and anti-particle ones. It is convenient to parse the problem and consider the arising matrix elements separately for particle-particle and particle-antiparticle transitions. Using the explicit matrix form of the unperturbed Hamiltonian $H_0$ from Eq.~(\ref{unpert_H}) and from now on suppressing the subscript $\mathbf{k}\lambda$ on the mode functions $\mathbf{f}_{\mathbf{k}\lambda}$ that signals its dependence on photon wave vector and polarization, we have:
\begin{equation}
(H_0+E_{\nu '})(\boldsymbol{\alpha}\cdot\mathbf{f})(H_0+E_\nu )=\left( \begin{array}{cc}
H_{ee} & H_{e\Bar{e}} \\ 
  H_{e\Bar{e}} &  H_{\Bar{e}\Bar{e}}
\end{array} \right) \label{HMatrix}
\end{equation}
The explicit forms of the matrix elements are:
\begin{align*}
H_{ee}&=(E_{\nu '}+m)(\boldsymbol{\sigma}\cdot\mathbf{f})(\boldsymbol{\sigma}\cdot\boldsymbol{\pi})+ (E_{\nu }+m)(\boldsymbol{\sigma}\cdot\boldsymbol{\pi})(\boldsymbol{\sigma}\cdot\mathbf{f}) \\
H_{e\Bar{e}}&=(\boldsymbol{\sigma}\cdot\boldsymbol{\pi})(\boldsymbol{\sigma}\cdot\mathbf{f})(\boldsymbol{\sigma}\cdot\boldsymbol{\pi})+ (E_{\nu '}-m)(E_{\nu }+m)(\boldsymbol{\sigma}\cdot\mathbf{f})\\
H_{\Bar{e}\Bar{e}} &= (E_{\nu }-m)(\boldsymbol{\sigma}\cdot\boldsymbol{\pi})(\boldsymbol{\sigma}\cdot\mathbf{f})+ (E_{\nu '}-m)(\boldsymbol{\sigma}\cdot\mathbf{f})(\boldsymbol{\sigma}\cdot\boldsymbol{\pi})\; ,
\end{align*}
where the momentum of the particle moving in the external magnetic field has been denoted by 
\begin{equation}
\bm{\pi} = \B{p}-e \B{A}_0\;. 
\end{equation}
We list $H_{\Bar{e}\Bar{e}}$ only for completeness; it is irrelevant for the energy shift of initial states that are particle rather than anti-particle states. 

In order to find magnetic moment contributions up to order $1/m^3$, we expand each term in Eq.~(\ref{PertMainEqn}) for large $m$ and extract all terms proportional to $\sigma_z B_0$,  while taking care to include, where necessary, the effects of the multipole expansion of the field, Eq.~(\ref{multipole}). When manipulating the above matrix elements one needs to remember that $\bm{\pi}$ contains a differential operator that does not commute with any quantity that is a function of position; its commutator with a function $g({\bf r)}$ is $[\pi_i,g({\bf r})] =-i\nabla_i g(r)$. With that in mind we use the algebra of the $\sigma$ matrices, $\sigma_i \sigma_j=\delta_{ij}+i\epsilon_{ijk}\sigma_k$, and decompose $H_{ee}$ along the lines of Eq.~(\ref{QOperator}) to rewrite it as
\begin{align}
H_{ee}&= \!\hat{b}_R^\dagger \beta_0\Big[ i (E_\nu +E_{\nu '}+2m)\minus{f}-(E_{\nu '}-E_\nu )\minus{(\boldsymbol{\sigma} \times  \mathbf{f})}\Big] \notag  \\
& -\hat{b}_R\beta_0 \Big[ i(E_\nu +E_{\nu '}+2m)\plus{f}-(E_{\nu '}-E_\nu )\plus{(\boldsymbol{\sigma} \times  \mathbf{f})}\Big] \notag \\
&+ (E_\nu +E_{\nu '}+2m)f_zp_z+(E_{\nu '}-E_\nu )(\boldsymbol{\sigma} \times  \mathbf{f})_z\notag \\
& + (E_{\nu '}+m) \boldsymbol{\sigma} \cdot ( \nabla\times \mathbf{f}) \;.
\label{Heeplusminus}
\end{align}
The expression we need to consider is
\begin{equation}
 e^2 \sum_{\nu ',s'} \sum_\lambda \int\!\! d^3k\frac{|\bra{\nu ,s}H_{ee}\ket{\nu ',s'}|^2}{4(E_\nu -E_{\nu '}-\omega)E_\nu (E_\nu +m)E_{\nu '}(E_{\nu '}+m)}
 \label{Hee}
\end{equation}
Carrying out the large $m$ expansion and discarding terms higher than order $1/m^3$, we find that none of the $\boldsymbol{\sigma} \times \mathbf{f}$ terms contribute to the energy shift to this order, as the denominator of Eq.~(\ref{Hee}) is already yielding terms of order $1/m^4$. The weak magnetic field expansion $B_0\rightarrow 0$ of the energy difference $E_\nu -E_{\nu '} -\omega$ generates factors of $eB_0/(m\omega)$. Thus the significant terms from the numerator have to be of order $m^2$. Of those, the terms in $f_{-}f_{-}^*$ and $f_{+}f_{+}^*$ and those involving $p_z^2$ are spin-independent, so do not contribute to the magnetic moment. This leaves three terms, whose matrix elements are proportional to
\begin{align}
&\bra{\nu,s} f_{-} \ket{\nu',s'}\bra{\nu',s'} \boldsymbol{\sigma} \cdot (\nabla \times \mathbf{f}^*)\ket{\nu, s}+\text{H.c.}\notag \\
&\bra{\nu,s} f_{+} \ket{\nu',s'}\bra{\nu',s'} \boldsymbol{\sigma} \cdot (\nabla \times \mathbf{f}^*)\ket{\nu, s}+\text{H.c.}\notag \\
&|\bra{\nu,s}\boldsymbol{\sigma} \cdot (\nabla \times \mathbf{f})\ket{\nu',s'}|^2+\text{H.c.} \label{ThreeTerms}
\end{align}
One might think that the first two terms cannot contribute, since the matrix elements on the left cannot change the spin $s$ and the matrix elements on the right at first sight appear not to be capable of changing $\nu$, but the latter is true only in the dipole approximation. Taking along the next-to-leading term in the multipole expansion of $\boldsymbol{\sigma} \cdot (\nabla \times \mathbf{f}^*)$, we see that they in fact do admit transitions between Landau levels. The contribution to the energy shift from transitions into the intermediate state with $\nu '=\nu -1$ is
\begin{widetext}
\begin{equation}
\Delta E_{\text{quad}}(\nu ' = \nu -1) = \frac{e^3B_0}{2m^3}\sum_\lambda \int d^3\B{k} 
\frac{i\beta_0}{\omega^2}\sqrt{\nu } \minus{f} \bra{\nu -1,s}(\mathbf{r}-\mathbf{r}_0)\cdot \nabla( \boldsymbol{\sigma} \cdot ( \nabla\times  \mathbf{f}^*))\ket{\nu,s } + \text{H.c.}\;,
\end{equation}
and the contribution from those with $\nu '=\nu +1$ is very similar. Substituting the matrix elements of the position operator from Eq.~(\ref{position_elements}), we see that the factors of $\beta_0$ cancel and this contribution is indeed of order $B_0$ in the external magnetic field. The sum of the $\nu '=\nu +1$ and $\nu '=\nu -1$ terms can be written as a set of second partial derivatives of mode functions. We can simplify the resulting expression if we take into account the orthogonality properties of the polarization vectors of the electromagnetic field (which we shall discuss in detail in the next section) and the fact that some terms are odd under the even integration over all photon modes and hence drop out.
We find for the total contribution from these quadrupole terms \footnote{Energy shifts depend on the state being spin-up or spin-down. Here and throughout this paper we abbreviate this dependence by writing energy shifts as proportional to the Pauli spin matrix $\sigma_z$ which should be understood as a shorthand for $\bra{s} \sigma_z \ket{s}$. }
\begin{equation}
\Delta E_{\text{quad}}=-\frac{e^3\sigma_zB_0}{4m^3} \sum_\lambda \int d^3\B{k}  \frac{1}{\omega^2} \Bigg( f_y\pd{^2f_y^*}{x^2} - f_y\pd{^2f_x^*}{x\partial y} \notag  +f_x\pd{^2f_x^*}{y^2}   -f_x\pd{^2f_y^* }{x \partial y} \Bigg) + \text{H.c.} 
\end{equation}
\end{widetext}
The third line of Eqs.~(\ref{ThreeTerms}) contributes for $s\neq s'$, and in this case the dipole approximation is enough to deliver all terms that are relevant for the magnetic moment to order $1/m^3$. This spin-dependent term contributes
\begin{equation}
\Delta E_{\text{spin}} = \frac{e^3B_0\sigma_z}{4m^3}\sum_\lambda\! \int\!\! d^3\B{k} 
\frac{1}{\omega^2}\left(|{(\nabla \times \mathbf{f})}_x|^2 + |{(\nabla \times \mathbf{f})}_y|^2 \right)
\end{equation}
The sum of quadrupole and spin-dependent contributions gives the particle-particle portion of the shift,
\begin{equation}
\Delta E_{ee} = \Delta E_{\text{quad}}+\Delta E_{\text{spin}}\;.
\end{equation}

We now turn our attention to the matrix element between initial particle and intermediate anti-particle states, $H_{e\Bar{e}}$. Disentangling a product of three $\sigma$ matrices is a slightly lengthier procedure than dealing with just two of them in $H_{ee}$, but the calculation runs along exactly the same lines: one repeatedly applies the algebra of the $\sigma$ matrices. In addition to the non-commutation of the momentum $\boldsymbol{\pi}$ with any function of position, one now also has to take note that the vector product $\boldsymbol{\pi}\times\boldsymbol{\pi}$ does not vanish but has a $z$ component $(\boldsymbol{\pi}\times\boldsymbol{\pi})_z=-2i\beta_0^2$, which comes about because the $x$ and $y$ components of $\boldsymbol{\pi}$ do not commute, i.e. $[\pi_x,\pi_y]=-2i\beta_0^2$. Then straightforward calculation yields
\begin{widetext}
\begin{align*} 
H_{e\Bar{e}}=&-2\beta_0^2(\bRd\bRd  \minus{f} \minus{\sigma} +b_R b_R \plus{f} \plus{\sigma} ) + \bRd\beta_0\left\{ 2ip_z(\minus{f} \sigma_z + \minus{\sigma} f_z) +
  i \minus{(\nabla \times \mathbf{f})} - \minus{[\boldsymbol{\sigma} \times (\nabla \times \mathbf{f}) ]} \right\}\\ 
  &+ b_R \beta_0\left\{ 2ip_z(\plus{f} \sigma_z + \plus{\sigma} f_z)
  -i \plus{(\nabla \times \mathbf{f})} + \plus{[\boldsymbol{\sigma} \times (\nabla \times \mathbf{f}) ]} \right\} 
   +2 \beta_0^2 \bRd \bR (\minus{f}\plus{\sigma} + \plus{f}\minus{\sigma})+ 2\beta_0^2 \plus{f} \minus{\sigma}  \\ &- (\bm{\sigma} \cdot \mathbf{f}) \boldsymbol{\pi}^2 + e \mathbf{f} \cdot {\mathbf{B}_0} + ie \boldsymbol{\sigma} \cdot (\mathbf{f} \times \mathbf{B}_0) + (\boldsymbol{\sigma} \cdot \mathbf{f})(E_{\nu '} -m)(E_\nu  + m) \end{align*}
The denominator of Eq.~(\ref{PertMainEqn}) now gives terms of order $1/m^5$, and non-relativistic expansions of the various energy terms supplement this with factors of $eB_0/m^2$. Therefore an obvious contribution to the energy shift of order $eB_0/m^3$ comes from the final term in $H_{e\Bar{e}}$ and whose square contributes with $m^4$ in the numerator. It gives:
\begin{equation}
\Delta E_{e\bar{e},1}=\frac{e^3B_0}{2m^3}\sum_{s'}\sum_\lambda\! \int\! d^3k\; (1+2\nu+s+s')| \bra{s}  \boldsymbol{\sigma} \cdot \boldsymbol{f} \ket{ s'}|^2
=\frac{e^3B_0\sigma_z}{2m^3}\sum_\lambda\! \int\! d^3k\; |f_z|^2\;.
\end{equation}
The only other possible contributions are those for which the factor of $1/m^5$ supplied by the denominator is multiplied by a numerator of order $\beta_0^2 m^2\equiv -eB_0m^2/2$. Such contributions arise from the cross multiplication of the last term in $H_{e\Bar{e}}$ with terms that carry $\beta_0^2$ or $eB_0$. Among those, the terms with $\bRd\bRd$ and $\bR\bR$ can be ruled out as they lead into intermediate states with $\nu'=\nu\pm 2$ for which the other factor, the matrix elements of the final term in $H_{e\Bar{e}}$, vanishes. Therefore, the only contribution we still need to consider comes from a term proportional to
$$
\sum_{s'}\bra{s} 2 \beta_0^2 \nu (\minus{f}\plus{\sigma} + \plus{f}\minus{\sigma})+ 2\beta_0^2 \plus{f} \minus{\sigma} + eB_0 (\bm{\sigma} \cdot \mathbf{f}) (2\nu+1) + e \mathbf{f} \cdot {\mathbf{B}_0} + ie \boldsymbol{\sigma} \cdot (\mathbf{f} \times \mathbf{B}_0)
\ket{s'} \bra{s'} \boldsymbol{\sigma} \cdot \mathbf{f}^*\ket{ s}+\text{H.c.}
$$
\end{widetext}
As nothing inside the matrix elements depends on $s'$, the sum of $\ket{s'}\bra{s'}$ gives the identity operator. Multiplying out all the $\sigma$ matrices and collecting terms proportional to $\sigma_z$ is then straightforward and gives:
\begin{equation*}
\Delta E_{e\bar{e},2}= -\frac{e^3B_0\sigma_z}{4m^3}\sum_\lambda\! \int d^3\B{k}\; |f_z|^2\;.
\end{equation*}
Here we have again taken into account the orthogonality properties of the electromagnetic polarization vectors, on account of which e.g.~$\sum_\lambda\! \int\! d^3\B{k}\; f_x f_y^*=0 $.
Going beyond the dipole approximation is not necessary here and does not generate any additional contributions from $H_{e\bar{e}}$ to the accuracy of the energy shift that interests us.
Thus the energy shift due to intermediate antiparticle states is:
\begin{equation}
\Delta E_{e\bar{e}} =\Delta E_{e\bar{e},1}+\Delta E_{e\bar{e},2}= \frac{e^3B_0\sigma_z}{4m^3 }\sum_\lambda\! \int d^3\B{k}\;  |f_z|^2 \;.
\end{equation}
From the total energy shift $\Delta E = \Delta E_{ee} + \Delta E_{e\bar{e}}$, we extract the following magnetic moment shift, via $\Delta E = -\Delta \mu_\perp \sigma_z B_0$,
\begin{widetext}
\begin{equation}
 \Delta \mu_\perp=-\frac{e^3}{4m^3}\sum_{\substack{\text{all}\\  \text{modes}}}\left\{{ |f_z|^2}+\frac{|{(\nabla \times \mathbf{f})}_x|^2}{\omega^2} + \frac{|{(\nabla \times \mathbf{f})}_y|^2}{\omega^2} +\frac{1}{\omega^2}\left(f_x\frac{\partial^2f_y^* }{\partial x \partial y}+ f_y\frac{\partial^2f_x^*}{\partial x\partial y} -f_y\frac{\partial^2f_y^*}{\partial x^2}  -f_x\frac{\partial^2f_x^*}{\partial y^2}     + \text{H.c.} \right)\right\} \label{ShiftInTermsofModeFunctions}
\end{equation}
\end{widetext}
This constitutes the first major result of this paper --- an expression that delivers the magnetic moment shift for an electron interacting with \emph{any} quantized electromagnetic field subjected to boundary conditions. We have presented the calculation with an externally applied magnetic field ${\bf B}_0$ that is perpendicular to the surface of the medium in mind, but the orientation of any surface has not actually entered the calculation at any stage. Hence we can quite easily find the corresponding shift for a magnetic field parallel to the surface by choosing the mode functions ${\bf f}({\bf r})$ such that they are solutions of the wave equation with a differently oriented surface of the dielectric, or, equivalently, cycle the Cartesian coordinates in Eq.~(\ref{ShiftInTermsofModeFunctions}). One can also apply the formula to the case of non-planar surfaces, provided one can convert the mode functions to Cartesian coordinates. 

We note that our derivation and the resulting Eq.~(\ref{ShiftInTermsofModeFunctions}) can of course not be used to calculate the anomalous magnetic moment in free space. While one could fudge an estimate by cutting off the integral over photon frequencies at $\omega\sim m$, which would give the correct order of magnitude $e^3/m$, a correct calculation would require the field quantization not just of the photon but also of the electron. As explained earlier, if we are interested only in the shift of the magnetic moment compared to its free-space value and if the interaction with the surface is wholly electromagnetic so that the electron is not subject to boundary conditions due to the presence of the surface, the electron propagator is not affected by the interaction with the surface and we can calculate the shift by using a quantum-optical approach and quantize only the photon field. 

\section{Electromagnetic mode functions}
\subsection{Nondispersive dielectric} \label{nondispmodes}
\begin{figure}
\includegraphics{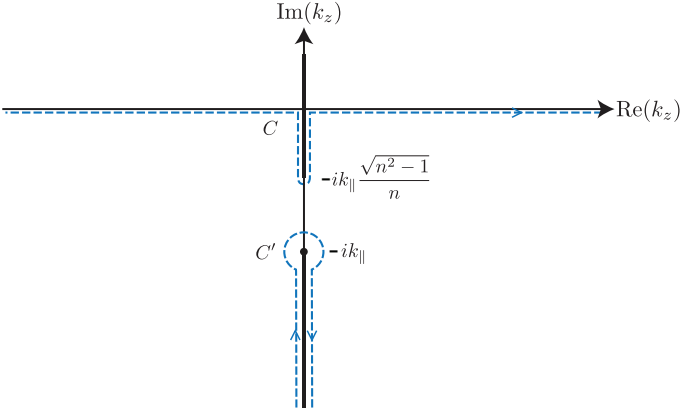}
\caption{\label{fig:contourdielectric} (Color online) Complex $k_z$ plane for nondispersive dielectric}
\end{figure}

We wish to determine the magnetic moment shift (\ref{ShiftInTermsofModeFunctions}) for a range of models for the material 
and start by considering a semi-infinite slab of non-magnetic, nondispersive material that fills the half-space $z>0$ as shown in Fig.~\ref{fig:HalfSpace}. The dielectric function is
\begin{equation}
\epsilon({\bf r}) = 1+\Theta(z)(n^2-1)
\label{epsilon_nondisp}
\end{equation}
where $n^2\geq 1$ is the index of refraction. 
Maxwell's equations in a dielectric but non-magnetic medium with the dielectric function $\epsilon({\bf r})$ give the following equation for the quantized vector field ${\bf A}_Q$,
\begin{equation}
\nabla\times(\nabla\times {\bf A}_Q)= -\epsilon({\bf r})\frac{\partial^2{\bf A}_Q}{\partial t^2}
\label{Maxwell_A}
\end{equation}
For a piecewise constant dielectric function, like the one in Eq.~(\ref{epsilon_nondisp}), the solutions of the above differential equation are found most efficiently by employing the generalized Coulomb gauge, defined by
\begin{equation}
\nabla \cdot [\mathbf{\epsilon(\mathbf{r})A}_Q]=0 \;,
\end{equation} 
so that, except directly at the interface $z=0$, Eq.~(\ref{Maxwell_A}) reduces to the standard wave equation with plane-wave solutions. These plane wave solutions are then patched together at the interface $z=0$ by imposing the continuity conditions that follow directly from Maxwell's equations,
\begin{eqnarray}
\epsilon(z) E_z(\B{r})|_{z=0^-} &= &\epsilon(z) E_z(\B{r})|_{z=0^+} \nonumber\\
\mathbf{E}_\parallel(\B{r}) |_{z=0^-} &=& \mathbf{E}_\parallel (\B{r}) |_{z=0^+}\label{continuity}\\
\mathbf{B}(\B{r}) |_{z=0^-} &= &\mathbf{B}(\B{r}) |_{z=0^+}\;.\nonumber
\end{eqnarray}
In the following we shall write wave vectors that exist on the vacuum side as $\mathbf{k}$, and those on the dielectric side as $\mathbf{k}^d$. A superscript $R$ will denote a reflected wave vector with a reversed $z$ component. It will be convenient to decompose wave vectors ${\bf k}$ into components parallel to the surface ($\mathbf{k_\parallel}$) and perpendicular to it ($k_z$). Modes will be labelled by their direction of incidence, incident momentum $\mathbf{k}$, and polarization $\lambda$, and will be broken down into incident, reflected and transmitted parts. Our notation is such that sgn$(k_z) =$ sgn$(k_z^d)$ on the real axis, which corresponds to connecting incident and transmitted waves. Modes that are incident from inside the dielectric may suffer total internal reflection at the interface, and thus be evanescent on the vacuum side. This means that there is a certain range of values for $k_z^d$ whose corresponding $k_z$ are pure imaginary. The mode functions are 
\begin{widetext}
\begin{align}
\B{f}_{k \lambda,\text{nondisp}}^{\text{left}} &= \frac{1}{(2\pi)^{3/2}}\frac{1}{\sqrt{2\omega}} \left\{\Theta(-z)[e^{i\mathbf{k}\cdot \mathbf{r}} \hat{\B{e}}_\lambda(\mathbf{k})+R^L_\lambda e^{i\mathbf{k}^R\cdot \mathbf{r}} \hat{\B{e}}_\lambda(\mathbf{k}^R)] + \Theta(z)T^L_\lambda e^{i\mathbf{k}^d\cdot \mathbf{r}} \hat{\B{e}}_\lambda(\mathbf{k}^d) \right\} \notag \\
\B{f}_{k\lambda,\text{nondisp}}^{\text{right}} &=\frac{1}{(2\pi)^{3/2}} \frac{1}{\sqrt{2\omega}} \frac{1}{n}\left\{\Theta(z)[e^{i\mathbf{k}^d\cdot \mathbf{r}} \hat{\B{e}}_\lambda(\mathbf{k}^d)+R^R_\lambda e^{i\mathbf{k}^{dR}\cdot \mathbf{r}} \hat{\B{e}}_\lambda(\mathbf{k}^{dR})] + \Theta(-z)T^R_\lambda e^{i\mathbf{k}\cdot \mathbf{r}} \hat{\B{e}}_\lambda(\mathbf{k}) \right\}   \label{NonDispModes}
\end{align}
\end{widetext}
where the polarization vectors $\hat{\B{e}}_\lambda (\mathbf{k})$ obey the gauge condition $\mathbf{k} \cdot \hat{\B{e}}_\lambda(\mathbf{k})=0$. A convenient choice is a decomposition into transverse electric (TE) and transverse magnetic (TM) modes through
\begin{eqnarray}
\hat{\mathbf{e}}_\text{TE}(\mathbf{k}) &=& \frac{1}{k_\parallel}\left( k_y,-k_x,0\right)\nonumber\\
\hat{\mathbf{e}}_\text{TM}(\mathbf{k}) &=& \frac{1}{k k_\parallel}\left( k_xk_z,k_yk_z,-k_\parallel^2 \right)\;.\label{pol_vecs}
\end{eqnarray}
The reflection and transmission amplitudes in the mode functions (\ref{NonDispModes}) are given by the Fresnel coefficients
\begin{align*}
R^L_\text{TE} &= \frac{k_z - k_z^d}{k_z + k_z^d}  &T^L_\text{TE} &= \frac{2k_z}{k_z + k_z^d} \\
R^L_\text{TM} &= \frac{n^2 k_z - k_z^d}{n^2 k_z + k_z^d}  &T^L_\text{TM} &= \frac{2n k_z}{n^2 k_z + k_z^d} \\
R^R_\lambda &= - R^L_\lambda  &T^R_\lambda &= \frac{k_z^d}{k_z}T_\lambda^L 
\end{align*}
with $z$ components of the wave vectors in vacuum and in the medium connected by the laws of refraction,
\begin{equation}
k_z^d = \sqrt{n^2(k_z^2+k_\parallel^2)-k_\parallel^2}\, .
\end{equation}
The modes are the same as in Refs.~\cite{EberleinRobaschik,PRDEberleinRobaschik,CarnigliaMandel} but with slightly different conventions concerning their normalization. 
They are normalized such that the radiation Hamiltonian is represented as a collection of harmonic oscillators,
\begin{align}
H_\text{rad} &= \frac{1}{2} \int d^3{\bf r} \left[ \epsilon(\mathbf{r})\dot{\mathbf{A}}_Q^2 + (\nabla \times \mathbf{A}_Q)^2  \right]\\
& =\sum_\lambda\int d^3{\bf k} \; \omega \left( a_{\B{k}\lambda}^\dagger a_{\B{k}\lambda} + \frac{1}{2} \right) \label{Norm}
\end{align}
The modes form an orthogonal and complete set, as explicitly shown in Refs.~\cite{CarnigliaMandel, Hammer}. That this must be so can easily be seen by writing Eq.~(\ref{Maxwell_A}) for the mode functions in the form \cite{GlauberLewenstein} 
\begin{equation}
\frac{1}{\sqrt{\epsilon({\bf r})}}\nabla \times \left[\nabla \times \frac{1}{\sqrt{\epsilon({\bf r})}} \sqrt{\epsilon({\bf r})}\mathbf{f}_{\B{k}\lambda }(\mathbf{r})\right] = \omega^2  \sqrt{\epsilon({\bf r})}\mathbf{f}_{\B{k}\lambda}(\mathbf{r})\;. \label{OrthFirstStepNonDisp}
\end{equation}
Evidently, this is an eigenvalue problem for a Hermitean operator acting on $\sqrt{\epsilon({\bf r})}\mathbf{f}_{\B{k}\lambda }$, which must therefore form a complete set of orthogonal functions satisfying
\begin{equation}
\int d^3\mathbf{r}\; \epsilon(\mathbf{r}) f^*_{\B{k}\lambda}(\mathbf{r})f_{\B{k}'\lambda'}(\mathbf{r}) = \frac{1}{2\omega}\delta_{\lambda \lambda'} \delta^{(3)}(\mathbf{k}'-\mathbf{k})\;.
\label{OrthogonalityNonDisp}
\end{equation}
Substituting these modes into Eq.~(\ref{ShiftInTermsofModeFunctions}), we find that the magnetic moment shift can be written in the form: 
\begin{widetext}
\begin{align}
 \Delta \mu_\perp  
 &= -\frac{e^3}{4m^3}\sum_{\vartheta=\pm 1,\lambda}\int d^2\mathbf{k}_\parallel \Bigg\{   \int_0^\infty dk_z \, g^\vartheta_\lambda(k_\parallel, k_z) [1+|R^L_\lambda|^2] +\frac{1}{n^2} \int_{-\infty}^{-\Gamma} dk_z^d g^\vartheta_\lambda(k_\parallel, k_z) |T^R_\sigma|^2 
  \notag \\ & \qquad \qquad + \vartheta\int_0^\infty dk_z \, g^\vartheta_\lambda(k_\parallel, k_z)   R^L_\lambda(e^{2ik_zz}+e^{-2ik_zz})+ \frac{\vartheta}{n^2} \int_{-\Gamma}^0 dk_z^d g^\vartheta_\lambda(k_\parallel, k_z) |T^R_\sigma|^2 e^{2ik_z z} \Bigg\} \label{FormofDeltaMu}
\end{align}
\end{widetext}
where the functions $g^\vartheta_\lambda(k_\parallel, k_z)$ are
\begin{align}
g_{\text{TE}}^+  &= \frac{k_\parallel^2}{k^3}\ ,  &g_{\text{TE}}^-  &= \frac{k_z^2}{2k^3}\ ,\nonumber \\
g_{\text{TM}}^+  &= \frac{2k_\parallel^2+k_z^2}{2k^3}\ ,  &g_{\text{TM}}^- & =0\;. \label{g_def}
\end{align}
The critical value of $k_z^d$, below which the modes are evanescent on the vacuum side, is $\Gamma = \sqrt{n^2-1}\;k_\parallel$. 
The functions $g^\vartheta_\lambda(k_\parallel, k_z)$ have a branch cut due to $k=\sqrt{k_z^2+k_\parallel^2}$ in their denominators, but are otherwise analytic in $k_z$.
We place this cut at $k_z =\pm ik_\parallel\ldots \pm i \infty$. The contributions from the various types of modes to Eq.~(\ref{FormofDeltaMu}) are easily identifiable: the integrals over $k_z^d=-\infty\ldots -\Gamma$ and $k_z=0\ldots\infty$ correspond to right- and left-incident travelling modes, respectively, while the integral over $k_z^d=-\Gamma\ldots 0$ corresponds to evanescent modes.  $k_z^d$ has branch points at $k_z = \pm i\Gamma/n$; we place the branch cut in between.  Using $ dk_z^d = n^2 (k_z  /k_z^d)\; dk_z$, we can manipulate the $k_z$ and $k_z^d$ integrals from the first line of Eq.~(\ref{FormofDeltaMu}), which are all independent of $z$, into
\begin{align}
\int_0^\infty dk_z \, g^\vartheta_\lambda(k_\parallel, k_z) &\left[1+|R^L_\lambda|^2 + \frac{k_z}{k_z^d}  |T^R_\lambda|^2 \right] \notag \\
&= 2\int_0^\infty dk_z \, g^\vartheta_\lambda(k_\parallel, k_z)\;, \label{RTRelation}
\end{align}
where the equality follows since $k_z$ and $k_z^d$ are here both real. The second line of Eq.~(\ref{FormofDeltaMu}) contains the $z$ dependent terms and can be written as:
\begin{eqnarray}
 \vartheta\int_0^\infty dk_z \, g^\vartheta_\lambda(k_\parallel, k_z)   R^L_\lambda(e^{2ik_zz}+e^{-2ik_zz}) \nonumber\\
  + \vartheta \int_{0}^{i\Gamma/n} dk_z \frac{k_z}{k_z^d} g^\vartheta_\lambda(k_\parallel, k_z) |T^R_\sigma|^2 e^{2ik_z z}\;. \label{TwoIntegrals}
\end{eqnarray}
We observe that for real $k_z^d$ and pure imaginary $k_z$ 
\begin{equation}
R^L_{\sigma} |_{ k_z^d = -K} - R^L_{\sigma}|_{k_z^d = K} = \frac{k_z}{k_z^d} T^R_\sigma T^{R*}_\sigma |_{ k_z^d = -K}\;,
\end{equation}
which permits us to combine the integrals (\ref{TwoIntegrals}) into one,
\begin{equation}
\vartheta \int_C dk_z \,  g^\vartheta_\lambda(k_\parallel, k_z)  R^L_\lambda e^{2ik_z z}\;.
\end{equation}
with the contour $C$ as shown in Fig.~\ref{fig:contourdielectric}.
Thus, we have for the magnetic moment shift,
\begin{align}
 \Delta \mu_\perp =-\frac{e^3}{4m^3}\sum_{\vartheta=\pm 1,\lambda} & \int d^2\mathbf{k}_\parallel \Bigg[  \int_C dk_z \, g^\vartheta_\lambda(k_\parallel, k_z)   
 R^L_\lambda e^{2ik_z z}\notag \\
 & + 2\int_0^{\infty} dk_z \,  g^\vartheta_\lambda(k_\parallel, k_z)  \Bigg]
\end{align}
The second term in the brackets can be left out since it does not depend on the position $z$ of the particle. It arises from the same electromagnetic field fluctuations but in vacuum. i.e. without any dielectric medium present, as is easy to see by taking the limit $n\rightarrow 1$ and hence $ R^L_\lambda\rightarrow 0$ in the above equation. Therefore, we subtract it as a free-space counterterm and obtain for the renormalized, position-dependent magnetic moment shift
\begin{equation}
 \Delta \mu_\perp = -\frac{e^3}{4m^3}\sum_{\vartheta=\pm 1,\lambda}\int d^2\mathbf{k}_\parallel \int_C dk_z \,\vartheta g^\vartheta_\lambda(k_\parallel, k_z)  R^L_\lambda e^{2ik_zz} \label{DeltaMuForm} \, .
\end{equation}
The integrand is analytic in the lower half of the complex $k_z$ plane, so we can freely deform the contour $C$ into that labelled as $C'$ in Fig.~\ref{fig:contourdielectric}. 
Carrying out the sum over $\vartheta$ and the angular part of the $\mathbf{k}_\parallel$ integration, we find the following explicit expressions for the magnetic moment shifts 
$\Delta \mu_\perp$ and $\Delta \mu_\parallel$ in magnetic fields perpendicular and parallel to the surface of the dielectric medium:
\begin{widetext}
\begin{align}
\Delta \mu_\perp &=-\frac{e^3}{32\pi^2m^3}  \int_0^\infty dk_\parallel \int_{C'}dk_z \frac{k_\parallel}{k^3} \left[ \left( 2k_\parallel^2-k_z^2  \right)R^L_\text{TE}+\left(2k_\parallel^2+k_z^2 \right)R^L_\text{TM}\right] e^{2ik_zz} \label{ShiftAfterContourPerp}\;,\\ 
\Delta \mu_\parallel &=-\frac{e^3}{32\pi^2m^3}  \int_0^\infty dk_\parallel \int_{C'}dk_z \frac{k_\parallel}{2k^3} \left[ \left( 3k_\parallel^2+2k_z^2  \right)R^L_\text{TE}+\left(3k_\parallel^2-2k_z^2 \right)R^L_\text{TM}\right] e^{2ik_zz}\;. 
\label{ShiftAfterContourPara}
\end{align}
\end{widetext}
The shift $\Delta \mu_\parallel$ for the case of the externally applied magnetic field directed parallel to the surface of the dielectric 
has been evaluated in exactly the same way as $\Delta \mu_\perp$, but by cycling the Cartesian indices in Eq.~(\ref{ShiftInTermsofModeFunctions}) so as to describe a differently oriented surface, as explained below Eq.~(\ref{ShiftInTermsofModeFunctions}).

The integrals in Eqs.~(\ref{ShiftAfterContourPerp}) and \eqref{ShiftAfterContourPara} can be calculated exactly. However, we postpone discussion of their calculation and results until after consideration of a plasma surface, in order to be able to compare and contrast the two models.

\subsection{Plasma Surface} \label{PlasmaModes}
We now consider the simplest model of a dispersive medium --- an undamped plasma of freely moving charge carriers. The dielectric function for a half-space is then
\begin{equation}
\epsilon({\bf r},\omega) = 1-\Theta(z)\frac{\omega_p^2}{\omega^2} \label{PlasmaDielectricFunction}
\end{equation}
where $\omega_p\equiv\sqrt{Ne^2/m_c}$ is the plasma frequency which characterises the material made up from charge carriers of effective mass $m_c$, charge $e$, and an average density $N$. 
As discussed in detail in Ref.~\cite{ElsonRitchie}, Maxwell's equations yield the equation to be satisfied by the quantized vector field, and thus by the mode functions, as
\begin{equation}
\nabla \times [\nabla \times \mathbf{f}_{\B{k}\lambda }(\mathbf{r})]+ \Theta(z)\omega_p^2\mathbf{f}_{\B{k}\lambda}(\mathbf{r}) = \omega^2  \mathbf{f}_{\B{k}\lambda}(\mathbf{r})\;. \label{OrthFirstStep}
\end{equation}
Using Coulomb gauge one can simplify this to the wave equation, with a shifted frequency inside the material, and its solutions are travelling and evanescent waves. The continuity of the parallel wave vector ${\bf k}_\parallel$ means that the relation between the $z$  component of the wave vector inside the material, $k_z^d$, and the one in vacuum, $k_z$, is now given by
\begin{equation}
k_z^{d2} = k_z^2 - \omega_p^2 \label{Plasmakzd}\;.
\end{equation}
As $\omega_p$ is real, it is possible for $k_z$ and $k_z^d$ to be simultaneously imaginary, i.e.~there exist solutions to the wave equation which fall away exponentially on both sides of the interface; these are surface plasmons. 
Eq.~(\ref{Plasmakzd}) also tells us that if $k_z^d$ is real then $k_z$ must also be real, so that one cannot have modes originating from within the medium that become evanescent on the vacuum side. Thus, we expect to reproduce some of the features of the perfect reflector model. However, for $k_z < \omega_p$ the wave vector inside the medium, $k_z^d$, is pure imaginary and therefore the modes are travelling waves on the vacuum side but evanescent inside the medium. So, there are three types of mode: TE, TM and surface plasmon (SP). The TE and TM modes are always travelling waves in vacuum and can be travelling or evanescent waves in the medium. They can be written in almost the same form as in Eq.~(\ref{NonDispModes}) for the nondispersive dielectric. In order to see what is different, we note that 
the left hand side of Eq.~(\ref{OrthFirstStep}) is a Hermitian operator acting on $\mathbf{f}_{\B{k}\lambda}$, which implies the orthonormality relation
\begin{equation}
\int d^3 \mathbf{r}  \, \B{f}^*_{\B{k}'\lambda'}(\mathbf{r})\B{f}_{\B{k}\lambda}(\mathbf{r}) = \frac{1}{2\omega} \delta_{\lambda\lambda'}\delta^{(3)}(\mathbf{k}'-\mathbf{k})\;.
\label{plasma_ortho}
\end{equation}
The difference between this and the orthonormality relation (\ref{OrthogonalityNonDisp}) for the nondispersive case, is a factor of $\epsilon(\mathbf{r})$ under the integral. However, this is compensated by the different relation between $k_z$ and $k_z^d$ which now gives $dk_z^d=(k_z/k_z^d)dk_z$ and therefore $\delta(k_z-k_z')=(k_z/k_z^d)\delta(k_z^d-{k_z^d}')$ in Eq.~(\ref{plasma_ortho}). We find that, apart from the different dielectric function to be used within the reflection and transmission coefficients, the only difference in the TE and TM modes for the plasma as compared to the nondispersive dielectric is an overall factor of $n$ in the right incident modes. Thus, the TE and TM modes for the plasma are obtained from the nondispersive dielectric modes in Eq.~(\ref{NonDispModes}) via the replacements
\begin{align*}
\B{f}_{\B{k}\lambda\text{plasma}}^{\text{left}} &= \B{f}_{\B{k}\lambda,\text{nondisp}}^{\text{left}}(n^2 \to \epsilon(\omega)) \\
\B{f}_{\B{k}\lambda,\text{plasma}}^{\text{right}} &= n \B{f}_{\B{k}\lambda,\text{nondisp}}^{\text{right}}(n^2 \to \epsilon(\omega)) \;.
\end{align*}

The SP mode is derived by considering modes that exponentially decay on both sides of the surface. Defining
\begin{align*}
\kappa &\equiv i k_z = \sqrt{k_\parallel^2 - \omega_\text{SP}^2} \\
\kappa^d &\equiv -i k_z^d = \sqrt{k_\parallel^2 - \epsilon(\omega_\text{SP}) \omega_\text{SP}^2}
\end{align*}
one finds the solutions of Eq.~(\ref{OrthFirstStep}) to be of the form
$\exp(\kappa z+i \mathbf{k}_\parallel \cdot \mathbf{r}_\parallel)$ for $z<0$ and $\exp(-\kappa^d z+i \mathbf{k}_\parallel \cdot \mathbf{r}_\parallel)$ for $z>0$, with the polarization vectors as in Eq.~(\ref{pol_vecs}). The continuity conditions (\ref{continuity}) then relate the mode amplitudes on both sides of the interface. For the TE polarization the resulting equations turn out to contradict each other whence there is no TE surface plasmon. For the TM polarization the combination of these continuity conditions yields the relation 
\begin{equation}
\kappa^d = -\epsilon(\omega_\text{SP}) \kappa\;, \label{DispRelationStart}
\end{equation}
which shows that surface plasmons can occur only for frequencies where $\epsilon(\omega)$ is negative and delivers the surface plasmon dispersion relation,
\begin{equation}
\omega_\text{SP}^2 = k_\parallel^2 +\frac{\omega_p^2}{2}-\sqrt{k_\parallel^4+\frac{\omega_p^4}{4}} \label{PlasmaDispRelation} \;.
\end{equation}
As before, we find normalization constants by requiring that for each of the TE, TM and SP modes the Hamiltonian reduces to that of a set of harmonic oscillators, Eq.~(\ref{Norm}).
For the SP mode the set of oscillators exists only for the TM polarization and the sum over modes is just a two-dimensional integral over ${\bf k}_\|$, as the dispersion relation, Eq.~(\ref{PlasmaDispRelation}), fixes $\omega$ at the surface plasmon frequency $\omega_\text{SP}$, which also fixes $k_z$. 

The mode function for the surface plasmon is specific to the plasma model; it reads \cite{BabikerBarton,ElsonRitchie}
\begin{align}
{\bf f}_{{\bf k},\text{SP}} &= \frac{1}{2\pi} \frac{1}{\sqrt{p(k_\|)}}\left[\Theta(-z)\left(\hat{\bf k}_\parallel -\frac{i k_\parallel}{\kappa} \hat{\bf z}\right) e^{i \mathbf{k}_\parallel \cdot \mathbf{r}_\parallel+ \kappa z}
\right.\nonumber\\ 
&\left.\ \ \ +\Theta(z)\left(\hat{\bf k}_\parallel +\frac{i k_\parallel}{\kappa^d} \hat{\bf z}\right) e^{i \mathbf{k}_\parallel \cdot \mathbf{r}_\parallel- \kappa^d z}\right]\, , \label{Modes}
\end{align}
with the norming function
\begin{equation}
p(k_\|)=\frac{\epsilon^4-1}{\epsilon^2 \sqrt{-(1+\epsilon)}}\;,\quad\epsilon\equiv\epsilon(\omega_\text{SP})\;.
\label{normingfunction}
\end{equation}
Substituting the mode functions into Eq.~(\ref{ShiftInTermsofModeFunctions}), we find that, similarly to the case of the nondispersive dielectric, the magnetic moment shift due to the TE and TM modes may be written as
\begin{widetext}
\begin{equation}
 \Delta \mu_\perp^\text{TE,TM} =  -\frac{e^3}{4m^3}\sum_{\vartheta=\pm 1,\lambda}\int d^2\mathbf{k}_\parallel \left\{ \int_0^\infty dk_z \, g^\vartheta_\lambda(k_\parallel, k_z) [1+|R^L_\lambda|^2 +\vartheta  R^L_\lambda(e^{2ik_zz}+e^{-2ik_zz})] + \vartheta\int_{-\infty}^0 dk_z^d g^\vartheta_\lambda(k_\parallel, k_z) |T^R_\lambda|^2\right\}
\end{equation}
\end{widetext}
where the functions $g^\vartheta_\lambda(k_\parallel, k_z)$ are the same as in Eq.~(\ref{g_def}) for the nondispersive dielectric. 

\begin{figure}[b]
\includegraphics{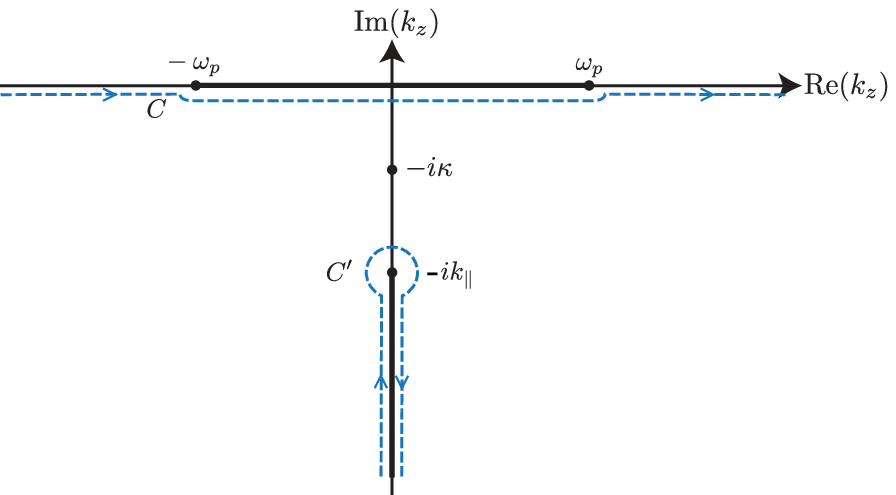}
\caption{\label{fig:contour} (Color online) Complex $k_z$ plane for plasma surface} \label{contourplasma}
\end{figure}

Again, the $z$ independent terms of this integral make up the free-space counterterm to be subtracted. In this case, these are all terms except the one proportional to $R^L_\lambda$. In order to see that these really give the free-space contribution, we use $dk_z^d=(k_z/k_z^d)dk_z$ to rewrite them as
\begin{align}
\int_0^{\omega_p} dk_z \, & g^\vartheta_\lambda(k_\parallel, k_z) \left(1+ |R^L_\lambda|^2 \right) \notag \\
& + \int^{\infty}_{\omega_p} dk_z  g^\vartheta_\lambda(k_\parallel, k_z)\left(1+|R^L_\lambda|^2+ \frac{k_z}{k_z^d}|T^L_\lambda|^2 \right).
\label{freespace}
\end{align}
Here $k_z$ is always real, as there are no evanescent modes on the vacuum side. However, $k_z^d$ is imaginary for $k_z<\omega_p$, which means that in the first integral $|R^L_\lambda |^2=1$. The second integral can be simplified by applying Eq.~(\ref{RTRelation}) as $k_z$ and $k_z^d$ are both real. Therefore, Eq.~(\ref{freespace}) is just a constant, independent of the properties of the material and the same as one would get in free space. Subtracting it as free-space counterterm from $\Delta \mu_\perp^\text{TE,TM}$, we arrive at the same expression as in Eq.~(\ref{DeltaMuForm}), except with the contour running straight along the real axis as shown in Fig.~ \ref{contourplasma}. The integrands (up to the explicit form of $\epsilon$ within $R^L_\lambda$) for the TE and TM modes in the plasma model are identical to those for the nondispersive dielectric.
Just as with the nondispersive case, we deform this contour into the lower half-plane. We note that for the plasma model the TM reflection coefficient has a pole at 
$-i\kappa=\sqrt{\omega_{sp}^2-k_\parallel^2}$. 
We are left with a contribution from the contour integral over $C'$ that is identical to Eqs.~(\ref{ShiftAfterContourPerp}) and (\ref{ShiftAfterContourPara}) (the corresponding quantities in the nondispersive case), and a contribution proportional to the residue of the TM reflection coefficient. Now we add the surface plasmon part, which is derived by substituting the surface plasmon mode functions, Eq.~(\ref{Modes}), into the expression (\ref{ShiftInTermsofModeFunctions}) for $\Delta\mu_\perp$. This gives the magnetic moment shift due to the surface plasmon as
\begin{align}
\Delta \mu_\perp^\text{SP} &= -\frac{e^3}{8\pi m^3}\int_0^\infty d k_\parallel k_\parallel \frac{2 k_\parallel^2-\kappa ^2}{ p(k_\parallel)   \kappa ^2} e^{2  \kappa z }, \label{SPOnly_perp}\\
\Delta \mu_\parallel^\text{SP} &= -\frac{e^3}{8\pi m^3}\int_0^\infty d k_\parallel k_\parallel \frac{3 k_\parallel^2+2\kappa ^2}{ 2p(k_\parallel)   \kappa ^2} e^{2  \kappa z } .\label{SPOnly_para}
\end{align}
Adding these to the results for $\Delta \mu_\perp^\text{TE,TM}$ we find that they exactly cancel the contributions from the residue part of the contour integration along $C'$, similarly to what has been found in the calculation of atomic energy level shifts near a plasma surface \cite{BabikerBarton}. Thus for the plasma surface the total magnetic moment shift, i.e.~the sum over TE, TM \emph{and} surface plasmons contributions, is given precisely by Eqs.~(\ref{ShiftAfterContourPerp}) and (\ref{ShiftAfterContourPara}). In other words, the magnetic moment shift for both the nondispersive dielectric surface and the plasma surface can be found from the \emph{same} integral simply by inserting the appropriate dielectric function into the reflection coefficients. The additional modes arising in the plasma model are automatically taken care of by the more complicated structure of the expression in the complex $k_z$ plane over which the contour is deformed, with the same end result. 

\subsection{Dispersive Dielectric}\label{DispDiel}
The next model we would like to consider is a dispersive dielectric. For this we need to move the pole in the dielectric function of a plasma away from zero frequency to a finite transverse optical phonon resonance $\omega_T$, which corresponds to the inclusion of a restoring force into the equation of motion for the electrons within the material \cite{BartonPlasma}. The dielectric function is then
\begin{equation}
\epsilon(\mathbf{r},\omega) = 1- \Theta(z) \frac{\omega_p^2}{\omega^2-\omega_T^2}. \label{DispersiveDielectricFunction}
\end{equation}
The dispersion relation for the surface polariton turns from Eq.~(\ref{PlasmaDispRelation}) into
\begin{equation}
\omega_{sp}^2 = k_\parallel^2 + \frac{1}{2}(\omega_p^2+ \omega_T^2) - \sqrt{k_\parallel^4 - k_\parallel^2\omega_T^2 + \frac{1}{4}(\omega_p^2+\omega_T^2)^2}\;. \label{NewDispRelation}
\end{equation}
The quantization of the electromagnetic field in terms of normal modes is now hindered by the fact that the field equation
\begin{equation}
\nabla \times [\nabla \times \mathbf{A}_Q(\mathbf{r},\omega)]+ \frac{\Theta(z)\:\omega_p^2}{1-\omega_T^2/\omega^2}\mathbf{A}_Q(\mathbf{r},\omega) = \omega^2  \mathbf{A}_Q(\mathbf{r},\omega) \label{field_dispdiel}
\end{equation}
cannot be written as a Hermitian eigenvalue problem, which would have guaranteed the orthogonality and completeness of the modes. 
Thus, for a first-principles derivation of the magnetic moment shift for this kind of surface one would need to include both dispersion and absorption into the model and construct a Huttner-Barnett-type field theory for the electromagnetic field interacting with the dielectric medium (for a suitable formulation see e.g. \cite{RJZarxiv}).
\begin{figure}
\includegraphics{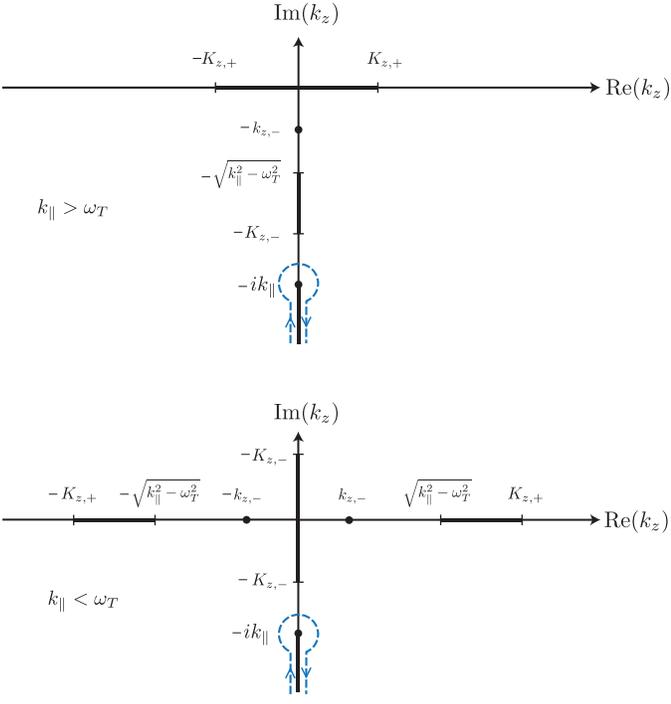}
\caption{(Color online) Complex $k_z$ plane for dispersive dielectric surface. } 
\label{fig:contourdispdiel}
\end{figure} 
 
An alternative approach \cite{REBthesis} is to use a Lifshitz-style method \cite{Lifshitz} and write the expectation values of squares of electromagnetic field operators, as in Eq.~(\ref{ShiftInTermsofModeFunctions}), in terms of a Green's tensor with an arbitrary permittivity $\epsilon({\bf r},\omega)$. 
This tensor turns out to depend only on the reflection coefficients of the surface \cite{MaradudinMills}, just as our formulae (\ref{ShiftAfterContourPerp}) and (\ref{ShiftAfterContourPara}) do. Thus, for an end result that can depend only on the surface's reflection coefficients, one necessarily gets the same expressions (\ref{ShiftAfterContourPerp}) and (\ref{ShiftAfterContourPara}), as before, for the magnetic moment shifts. Likewise, the Feynman propagator of the Huttner-Barnett field theory constructed in Refs.~\cite{RJZarxiv,RJZanisotropic} depends only on the same reflection coefficients, thus leading to the same conclusion that such a field-theoretical approach necessarily reproduces Eqs.~(\ref{ShiftAfterContourPerp}) and (\ref{ShiftAfterContourPara}) for the magnetic moment shift.

The continued validity of Eqs.~(\ref{ShiftAfterContourPerp}) and (\ref{ShiftAfterContourPara}) for a dispersive dielectric is facilitated by the fact the integration path $C'$ in the complex $k_z$ plane is not affected by the introduction of the transverse optical resonance $\omega_T$.
The complex $k_z$ plane for a dispersive dielectric model is shown in Fig.~\ref{fig:contourdispdiel} with the abbreviations
\begin{eqnarray}
K_{z,\pm} = &&\frac{1}{\sqrt{2}}\left\{\omega _p^2+\omega _T^2 -k_\parallel^2 \right.\nonumber\\
&&\left.\pm \sqrt{k_\parallel^4+2 k_\parallel^2 \left(\omega _p^2-\omega _T^2\right) +\left(\omega _p^2+\omega _T^2\right)^2} \right\}^{1/2}\nonumber
\end{eqnarray}
and 
$$k_{z,\pm} =\sqrt{\frac{1}{2}(\omega _p^2+\omega _T^2)\pm \sqrt{ k_\parallel^4- k_\parallel^2 \omega _T^2+\frac{1}{4}
\left(\omega _p^2+\omega _T^2\right)^2}}. $$
While the structure of the complex $k_z$ plane is considerably more complicated for the dispersive dielectric model as opposed to the plasma model, none of the additional cuts or poles interfere with the integration path $C'$, and thus Eqs.~(\ref{ShiftAfterContourPerp}) and (\ref{ShiftAfterContourPara}) can be applied as they are.

\section{Evaluating the magnetic moment shift} \label{Integral}

In order to calculate the magnetic moment shifts (\ref{ShiftAfterContourPerp}) and (\ref{ShiftAfterContourPara}) for the various models of the surface, we divide up the integrals into the contribution from the two straight lines on either side of the cut, $k_z = (-i\infty,-ik_\parallel]$ and $k_z =[-ik_\parallel, -i\infty)$, and the contribution from the small circle around $k_z = -ik_\parallel$. The integral around this small circle proves awkward to evaluate, so we subtract that point and consider it separately. This separate integral turns out to be elementary; the result appears below as the terms outside the integrals. We write the full results in terms of the complex frequency $\xi  = i\omega$ and of $\eta = k_z/\omega$, which is the cosine of the complex angle of incidence. For isotropic media, this means that the dielectric function $\epsilon$ is a function only of $\xi$ and the material parameters of the surface, like $n$, $\omega_p$, and $\omega_T$. The magnetic moment shifts for the magnetic field ${\bf B}_0$ perpendicular and parallel to the interface are
\begin{widetext}
\begin{align}
\Delta \mu_\perp &= \frac{e^3}{16 \pi^2  m^3} \left\{\int_0^\infty d\xi \xi \int^\infty_1 d\eta \left[ \left( 3\eta^2-2\right)R^L_\text{TE}+\left(\eta^2-2 \right)\left(R^L_\text{TM}-\frac{\epsilon(0)-1}{\epsilon(0)+1}\right)\right] e^{2\xi \eta z} -  \frac{\epsilon(0)-1}{\epsilon(0)+1} \frac{3}{4  z^2 } \right\} \label{FullResultPerp} \\
\Delta \mu_\parallel &= \frac{e^3}{16 \pi^2  m^3} \left\{\frac{1}{2}\int_0^\infty d\xi \xi \int^\infty_1 d\eta \left[ \left( \eta^2-3\right)R^L_\text{TE}+\left(5\eta^2-3 \right)\left(R^L_\text{TM}-\frac{\epsilon(0)-1}{\epsilon(0)+1}\right)\right] e^{2\xi \eta z} -  \frac{\epsilon(0)-1}{\epsilon(0)+1} \frac{1}{ z^2 } \right\} \label{FullResultPara}
\end{align}
\begin{equation}
\text{with} \qquad R^L_\text{TE} = \frac{\eta -\sqrt{(\epsilon(\xi)-1)+\eta^2}}{\eta+\sqrt{(\epsilon(\xi)-1)+\eta^2}} 
\quad, \quad  R^L_\text{TM} = \frac{\eta \epsilon(\xi)-\sqrt{(\epsilon(\xi)-1)+\eta^2}}{\eta\epsilon(\xi)+\sqrt{(\epsilon(\xi)-1)+\eta^2}}\ .
\label{Rcoeffs}
\end{equation}
\end{widetext}
We note that there is no need for the subtraction from $R_\text{TE}^L$ of $R_\text{TE}^L(k_z\to -ik_\parallel)$ because that is zero. 

As shown later, for certain dielectric functions there may arise problems due to non-commutation between limits of physical parameters and the limit the $k_z\to -ik_\parallel$, which corresponds to the static limit $\omega\rightarrow 0$, and this will prove to be crucial to the analysis and comparison of the results for the various models of the surface. 
The subtraction of the point at $k_z \to -ik_\parallel$ is, in these variables, the subtraction of the point $\{\xi\to0, \eta \to \infty \}$. Taking this two-dimensional limit risks the obvious pitfall of potential non-commutation of the $\xi$ and $\eta$ limits. This complication arises with the TE part of the integrals for the plasma surface. We emphasize that this is not an artefact of using these variables --- even in an alternative formulation with $\{\xi\to0, \eta \to \infty \}$ expressed as a single-variable limit, the issue manifests itself, though in a different way. Fortunately, this one problematic case can be dealt with differently, as detailed in Appendix \ref{TEPlasmaAppendix}. Thus, the above formulae deliver the magnetic moment shift for a specified dielectric function $\epsilon({\bf r},\omega)$, as long as the limits $ \xi\to0$ and  $\eta \to \infty$ of the reflection coefficients commute. 

\begin{figure}
\includegraphics{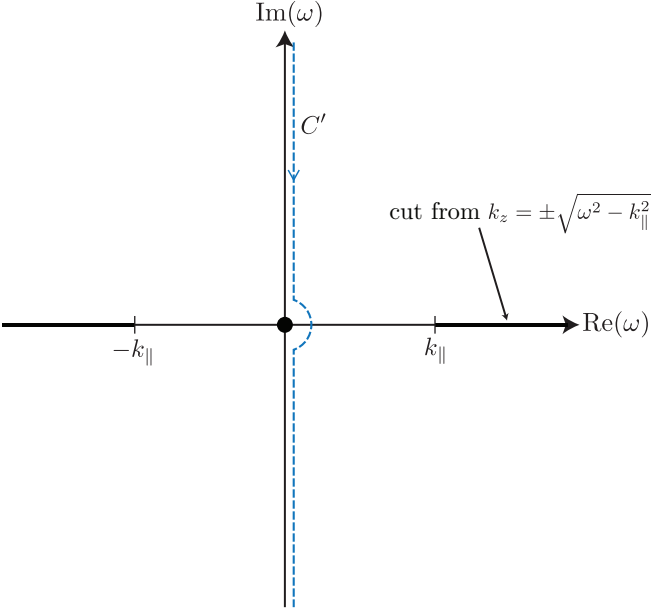}
\caption{(Color online) Integration path $C'$ in the complex $\omega$ plane. } 
\label{fig:omega_plane}
\end{figure}

If one carries out the corresponding calculation in either a Huttner-Barnett field theoretical approach, as in Refs.~\cite{RJZarxiv,RJZanisotropic}, or a noise-current approach \cite{REBthesis}, then the magnetic moment shift is more naturally expressed not as an integral over $(k_z,k_\parallel)$, as in Eqs.~(\ref{ShiftAfterContourPerp}) and (\ref{ShiftAfterContourPara}), but as one over $(\omega,k_\parallel)$. Hence, for ease of comparison with such approaches, we now proceed to derive alternative expressions for the magnetic moment shifts.
Carefully considering the complex variable transformation from $k_z$ to $\omega$ one can transform Eqs.~(\ref{ShiftAfterContourPerp}) and (\ref{ShiftAfterContourPara}) to
\begin{widetext}
\begin{align}
\Delta \mu_\perp &=-\frac{e^3}{32\pi^2m^3}  \int_0^\infty dk_\parallel \int_{C'}d\omega\; \frac{k_\parallel}{k_z\omega^2} \left[ \left( 3k_\parallel^2-\omega^2  \right)R^L_\text{TE}+\left(k_\parallel^2+\omega^2 \right)R^L_\text{TM}\right] e^{2ik_zz} \label{OmegaShiftPerp}\;,\\ 
\Delta \mu_\parallel &=-\frac{e^3}{32\pi^2m^3}  \int_0^\infty dk_\parallel \int_{C'}d\omega\; \frac{k_\parallel}{2k_z\omega^2} \left[ \left( k_\parallel^2+2\omega^2  \right)R^L_\text{TE}+\left(5k_\parallel^2-2\omega^2 \right)R^L_\text{TM}\right] e^{2ik_zz}\;, 
\label{OmegaShiftPara}
\end{align}
\end{widetext}
with the transformed path $C'$ as shown in Fig.~\ref{fig:omega_plane} and $k_z=-i\sqrt{k_\parallel^2-\omega^2}$ for $|\omega|>k_\parallel$ along $C'$.
These formulae can be simplified by subtracting the $1/\omega^2$ term of the integrand's Laurent expansion around $\omega=0$ and treating it separately, when it integrates to zero along $C'$. Then, using the fact that the integrands are even in $\omega$ and re-writing the integrals in terms of the complex frequency $\xi=i\omega$, we obtain
\begin{widetext}
\begin{align}
\Delta \mu_\perp &= \frac{e^3}{16 \pi^2  m^3} \int_0^\infty dk_\parallel \int_0^\infty d\xi\; \frac{k_\parallel}{\xi^2} 
\left\{\frac{e^{2\sqrt{k_\parallel^2+\xi^2}z}}{\sqrt{k_\parallel^2+\xi^2}}\left[ \left( 3k_\parallel^2+\xi^2  \right)R^L_\text{TE}+\left(k_\parallel^2-\xi^2 \right)R^L_\text{TM}\right] -e^{2k_\parallel z}k_\parallel R^L_\text{TM}(0)\right\}\label{OmegaResultPerp} \\
\Delta \mu_\parallel &= \frac{e^3}{16 \pi^2  m^3} \int_0^\infty dk_\parallel \int_0^\infty d\xi\; \frac{k_\parallel}{2\xi^2} 
\left\{\frac{e^{2\sqrt{k_\parallel^2+\xi^2}z}}{\sqrt{k_\parallel^2+\xi^2}}\left[ \left( k_\parallel^2-2\xi^2  \right)R^L_\text{TE}+\left(5k_\parallel^2+2\xi^2 \right)R^L_\text{TM}\right] - 5e^{2k_\parallel z}k_\parallel R^L_\text{TM}(0) \right\}\label{OmegaResultPara}
\end{align}
\end{widetext}
with
$$
R^L_\text{TE} = \frac{\sqrt{\xi^2+k_\parallel^2} -\sqrt{\epsilon(\xi)\xi^2+k_\parallel^2}}{\sqrt{\xi^2+k_\parallel^2} +\sqrt{\epsilon(\xi)\xi^2+k_\parallel^2}}\ ,
$$
$$
R^L_\text{TM} = \frac{\epsilon(\xi)\sqrt{\xi^2+k_\parallel^2} -\sqrt{\epsilon(\xi)\xi^2+k_\parallel^2}}{\epsilon(\xi)\sqrt{\xi^2+k_\parallel^2} +\sqrt{\epsilon(\xi)\xi^2+k_\parallel^2}}\ ,
$$
and $R^L_\text{TM}(0)\equiv R^L_\text{TM}(\xi=0)$. This result can also be obtained from a noise-current approach \cite{REBthesis}.

The results in Eqs.~(\ref{OmegaResultPerp}) and (\ref{OmegaResultPara}) are completely equivalent to those in Eqs.~(\ref{FullResultPerp}) and (\ref{FullResultPara}). Which ones are preferable depends on the particular model, but generally speaking the difficulty in evaluating either is about the same for most dielectric functions $\epsilon(\xi=i\omega)$.

\section{Results} \label{Results}

\subsection{Non-Dispersive}
Inserting the dielectric function $\epsilon(\omega) = n^2$ into Eqs.~(\ref{FullResultPerp})--(\ref{Rcoeffs}), one can evaluate the integrals exactly. The results for the magnetic moment shifts in a magnetic field perpendicular and parallel to the surface are
\begin{widetext}
\begin{align}
\Delta \mu_\perp =&-\frac{e^3}{32\pi ^2m^3 z^2}\frac{1}{\left(n^4-1\right)^{3/2} }\Bigg[\sqrt{n^4-1} (5-2 n+n^2-2 n^3-3 n^4+n^5) \notag \\ 
&-n^4 \sqrt{n^2-1} \left(1+2 n^2\right) \text{arctanh}\left(\frac{(n-1) \sqrt{1+n^2}}{1+(n-1) n}\right)+2 \left(n^2-1\right) \left(1+n^2\right)^{5/2} \text{ln}\left(n+\sqrt{n^2-1}\right)\Bigg] \label{DeltaMuPerp}
\end{align}
\begin{align}
\Delta \mu_\parallel =&-\frac{e^3}{192\pi ^2m^3 z^2}\frac{1}{\left(n^4-1\right)^{3/2} }\Bigg[\sqrt{n^4-1}(26-9 n+8 n^2-23 n^3-3 n^4+n^5)\notag \\
&+3 n^4 \sqrt{n^2-1} \left(2-3 n^2\right)  \text{arctanh}\left(\frac{(n-1) \sqrt{1+n^2}}{1+(n-1) n}\right)+9 \left(n^2-1\right) \left(1+n^2\right)^{5/2} \text{ln}\left(n+\sqrt{n^2-1}\right)\Bigg]
\label{DeltaMuPara}
\end{align}
\end{widetext}
If we expand these shifts in a series for large values of the refractive index $n$ we obtain
\begin{equation}
\Delta\mu_\perp = -\frac{e^2}{4\pi}\frac{e}{2 m} \left(\frac{n}{4 \pi m^2 z^2}-\frac{1}{4 \pi m^2 z^2}+\mathcal{O}(1/n)\right) \label{Largennondisp}
\end{equation}
\begin{equation}
\Delta\mu_\parallel =-\frac{e^2}{4\pi}\frac{e}{2 m} \left(\frac{n}{24 \pi m^2 z^2}+\frac{1}{4 \pi m^2 z^2}+\mathcal{O}(1/n)\right)\;, \label{LargennondispRotated}
\end{equation}
which shows that the perfect-reflector limit does not exist and the shift diverges for $n\rightarrow\infty$. This is evidently unphysical, as it implies that the magnetic moment could be increased arbitrarily by increasing the refractive index $n$ of the surface. However, as we shall show below, this is in fact a misconception and it is that model of a dispersionless medium which is unphysical. Detailed comparison with the results for the magnetic moment shifts near a perfect reflector (cf. Eq. (7.12) of \cite{BartonFawcett}) reveals the rather curious fact that the next-to-leading terms independent of $n$  in Eqs. (\ref{Largennondisp}) and (\ref{LargennondispRotated}), if taken on their own, do in fact reproduce the results of the perfect-reflector case. It is easy to check that this is not a simple calculational error: if one takes the limit $n\rightarrow\infty$ in the reflection coefficients (\ref{Rcoeffs}) first and evaluates the equivalent of integrals (\ref{FullResultPerp}) and (\ref{FullResultPara}) afterwards (as shown in Appendix \ref{PerfectMirrorCalc}), one reproduces the results of Ref.~\cite{BartonFawcett}.
The calculation also reveals the mathematical origin of the discrepancy: the TE reflection coefficient differs depending on whether either the perfect reflector limit $n\to\infty$ or the static limit $\{\xi\to0, \eta \to \infty \}$, or $k_z \to -ik_\parallel$, is taken first. This indicates that the correct magnetic moment shift close to a specific material can only be obtained if the model chosen for the surface correctly reproduces the true low frequency behaviour of the dielectric susceptibility of the material. We shall elaborate on this point later on.

\subsection{Plasma Surface}
\begin{figure}[b]
\includegraphics[width = 8.7cm]{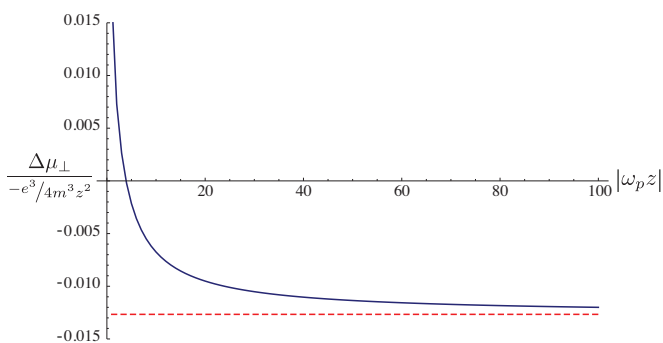}
\caption{\label{fig:plasmaresults}(Color online)  Comparison of plasma (solid line) and perfect reflector results (dashed line). (Both axes are in dimensionless natural units.)}
\end{figure}

Inserting the dielectric function (\ref{PlasmaDielectricFunction}) into the reflection coefficients (\ref{Rcoeffs}), we find that the $\eta\to\infty$ and $\xi\to0$ limits of $R_\text{TE}^L$ do not commute. This means we cannot use the TE parts of Eqs.~(\ref{FullResultPerp}) and (\ref{FullResultPara}) for this particular model. However, for the TE parts of these integrals we can go back to the stage before we deformed the contour and carry out the integration along the path $C$, because the TE reflection coefficients are very simple so that the integrals are unproblematic to calculate. As illustrated in Fig.~\ref{fig:contour}, the contour $C$ before deformation runs straight along the $k_z$ axis, passing under the cut. We show in Appendix \ref{TEPlasmaAppendix} how to calculate the integrals along the path $C$ for the TE contributions. The contribution from the TM modes can be evaluated from Eqs.~(\ref{FullResultPerp}) and (\ref{FullResultPara}). To calculate the integrals we first replace the integration over $\eta$ by one over $\kappa=\xi\eta$. In the resulting two-dimensional integral we can then change the order of integration and go from $\int_0^\infty d\xi\int_\xi^\infty d\kappa$ to $\int_0^\infty d\kappa\int_0^\kappa d\xi$. The integration over $\xi$ is elementary, and we are left with just a one-dimensional integral over $\kappa$. Simplifying the latter by scaling $\kappa$ to $s=\kappa/\omega_p$ we obtain
\begin{widetext}
\begin{align}
\Delta \mu_{\perp,\text{TM}} &= \frac{e^3}{16 \pi^2  m^3} \Bigg\{- \frac{3}{4  z^2 }+2\omega_p^2\int_0^\infty ds \; e^{2 s\omega_p z}\frac{ 1+t^2(s)}{t^2(s) \left[2+t^2(s)\right]^{3/2}}{ \left[2 t(s)-\left(1+2 t^2(s)\right)\text{arccot}(t(s))\right]}\Bigg\}, \label{plasmaTMPerp} \\
\Delta \mu_{\parallel,\text{TM}} &=  \frac{e^3}{16 \pi^2  m^3} \Bigg\{- \frac{1}{  z^2 }+\omega_p^2\int_0^\infty ds \; e^{2 s\omega_p z}\frac{ 1+t^2(s)}{t^2(s) \left[2+t^2(s)\right]^{3/2}}{ \left[3 t(s)-\left(5+3 t^2(s)\right)\text{arccot}(t(s))\right]}\Bigg\},\label{plasmaTMPara}
\end{align}
\end{widetext}
with the abbreviation
\begin{equation}
t(s) \equiv \sqrt{\sqrt{1+\frac{1}{s^2}}-1} \; .
\end{equation}
Note that we have included the constant terms from Eqs.~(\ref{FullResultPerp}) and (\ref{FullResultPara}) since these originate from the TM reflection coefficient. The remaining integral over $s$ is sufficiently complicated that it is best done numerically, which is easy since the exponential ensures fast convergence. We add to this the TE contribution calculated in Appendix \ref{TEPlasmaAppendix} and show the full results in Fig.~\ref{fig:plasmaresults}.

The $|\omega_p z| \to\infty$ asymptotics of Eqs.~(\ref{plasmaTMPerp}) and (\ref{plasmaTMPara}) is straightforward to work out by applying Watson's lemma. One finds that the $s$ integrals give terms of order $1/(\omega_pz^3)$ and are thus negligible compared to the $1/z^2$ leading-order contributions in the first summands and from the TE part (worked out in App.~\ref{TEPlasmaAppendix}). 
The $|\omega_p z| \to\infty$ limit of the overall result reproduces the perfect reflector result (and the $n$ independent terms in Eqs.~(\ref{Largennondisp}) and (\ref{LargennondispRotated})), as shown in Appendix \ref{TEPlasmaAppendix}. This makes sense physically, as the lack of evanescent modes on the vacuum side means that for $\omega_p\to\infty$ the plasma model becomes equivalent to the perfect reflector, thus reproducing its results.

By contrast, the plasma model does not reproduce the results of the nondispersive dielectric for large $n$. The consideration of a dispersive dielectric model in the following section will shed further light onto this issue and point to the origin of this discrepancy in the very different low-frequency behaviour of the electromagnetic response of conductors and insulators.

We also note that for the plasma model the $\omega_p\to 0$ and $\eta\to\infty$ limits of the TM reflection coefficient do not commute, causing the magnetic moment shift to diverge as $\omega_p \to 0$, when it should clearly be zero at this point. This arises because the static limit of the dielectric constant is ill-defined for $\omega_p\to 0$ and has already been discussed in detail in the context of the mass shift of an electron near a plasma surface \citep{massshift}. Of course, if we take $\omega_p \to 0$ before carrying out any of the integrals over photon wave vectors, there is no such problem and the shift vanishes as expected.

We have shown above that the shift for the nondispersive dielectric has a distance dependence of $1/z^2$ for all $z$ and all $n$, [cf.~Eqs.~(\ref{DeltaMuPerp}) and (\ref{DeltaMuPara})]. 
However, for the plasma model we find that for small distances, i.e.~small $|\omega_p z|$, the shift varies as $1/z^3$. 
The TE part of the shift contributes only logarithmically at small $|\omega_p z|$, as shown in Eq.~(\ref{ITESmallwpz}). So, the short-distance asymptotics of the shift is dominated by the TM parts of Eqs.~(\ref{FullResultPerp}) and (\ref{FullResultPara}), which we have already simplified into Eqs.~\eqref{plasmaTMPerp} and \eqref{plasmaTMPara}. To find the small $|\omega_p z|$ asymptotics, we scale the integration variable $s$ to a new variable equalling $s\omega_pz$ and then expand for small $|\omega_p z|$. The resulting series may then be integrated term by term and gives
\begin{align}
\Delta \mu_\perp (|\omega_p z| \ll1) &= \frac{e^3}{4m^3z^2}\left\{\frac{1}{16  \sqrt{2}  \pi }\frac{1}{ \omega_p z} + \mathcal{O}(\omega_p z)\right\}\label{SmallzperpPlasma} \\
\Delta \mu_\parallel (|\omega_p z| \ll1) &= \frac{e^3}{4m^3z^2}\left\{ \frac{5}{32  \sqrt{2} \pi  }\frac{1}{\omega_p z}  + \mathcal{O}(\omega_p z)\right\}\label{SmallzparaPlasma}
\end{align} 
The leading $1/z^3$ dependence seen here arises because at small distances the interaction between the electron and the surface is dominated by electrostatic interaction of the electron with the surface plasmon. Therefore one should be able to derive the $1/z^3$ term by considering the surface plasmon part of the mode functions (\ref{Modes}) alone. To this end, using the dielectric function (\ref{PlasmaDielectricFunction}) and the dispersion relation (\ref{PlasmaDispRelation}), we write the norming function $p(k_\parallel)$ [Eq.~(\ref{normingfunction})] and the imaginary $z$ component of the wave vector $\kappa\equiv ik_z$ as functions of $k_\parallel$ and $\omega_p$ and then substitute them into Eqs.~\eqref{SPOnly_perp} and (\ref{SPOnly_para}). Changing variables such that the $z$ dependence is taken out of the exponential and expanding for small $|\omega_p z|$, we find the leading term for the perpendicular case given by the trivial integral,
\begin{equation}
\frac{e^3}{16m^3 \sqrt{2} \pi  z^3 \omega_p }\int_0^\infty d x \, x ^2 e^{-2 x } =- \frac{e^3 }{64 m^3 \sqrt{2} \pi  z^3 \omega_p },
\end{equation}
which is in agreement with the $1/z^3$ term in Eq.~\eqref{SmallzperpPlasma}. For the parallel case the corresponding calculation reproduces Eq.~\eqref{SmallzparaPlasma}, as expected.
Thus the $1/z^3$ dependence of the shift at short distances does indeed originate entirely from the surface plasmon part of the mode functions, as observed in Ref.~\cite{BabikerPlasma} for the magnetic energy shift of a neutral atom interacting with a plasma surface. 

For large distances, the dependence of the magnetic moment shift remains $1/z^2$, as shown in Appendix \ref{TEPlasmaAppendix}. Since the inverse of the plasma frequency $\omega_p$ is the only length scale in the plasma model large and small distance regimes are defined by $|\omega_p z| \gg 1$ and $|\omega_p z| \ll 1$, respectively, and hence to speak of large distances is the same as speaking of large $\omega_p$.

\subsection{Dispersive Dielectric}

A dispersive dielectric has features in common with the nondispersive model, e.g.~that modes originating from within the material can be evanescent on the vacuum side. Likewise the TE reflection coefficient exhibits the same kind of problem with non-commuting limits: the limit of a large dielectric response, which is now described by $\omega_p \to \infty$, is not interchangeable with the static limit $\omega \to 0$. Thus for large $\epsilon$ we expect the dispersive dielectric to give similar results to the nondispersive dielectric.

\begin{figure}
\includegraphics[width=85mm]{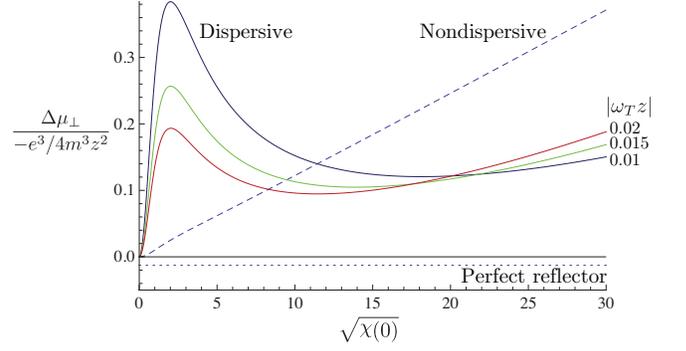}
\caption{\label{fig:graph}(Color online)  Magnetic moment shift for dispersive and nondispersive dielectric models as a function of static susceptibility, for the case of the magnetic field ${\bf B}_0$ perpendicular to the surface and $|\omega_T z|=\{0.01, 0.015, 0.02$.\}}
\end{figure}

We evaluate Eqs.~\eqref{FullResultPerp} and \eqref{FullResultPara} numerically, and find a peak in the magnetic moment shift relative to the perfect-reflector result. To facilitate the discussion of this peak and the comparison of different models, we now choose to write the dielectric function in terms of the static limit of the dielectric susceptibility, 
$$\chi(0) = \epsilon(0)-1=\omega_p^2/\omega_T^2\;.$$ 
We find peaks in $\Delta\mu_\perp$ and $\Delta\mu_\parallel$ at $\sqrt{\chi(0)} \approx 2$, with the height of the peak being inversely proportional to $\omega_T z$, as shown in Fig.~\ref{fig:graph} for the case where the external magnetic field ${\bf B}_0$ is perpendicular to the interface. We also plot the corresponding shift for the nondispersive case, where $\chi(0)_{\text{nondisp}}=n^2-1$. If we were to continue the plot to very large values of $\chi(0)$, the graphs for the two models would very slowly converge into one linearly-rising line. By contrast, the result for the perfect reflector, also shown in Fig.~\ref{fig:graph}, is much smaller and has the opposite sign.

The peak appears if the choice of parameters is such that $|\omega_T z|\lesssim 0.07$ for ${\bf B}_0$ perpendicular, and $\lesssim 0.25$ for ${\bf B}_0$ parallel to the surface. For smaller values of $|\omega_T z|$, the peak moves closer to $\sqrt{\chi(0)}\equiv\omega_p/\omega_T  \approx 2$, and increases in height. To gauge the enhancement that dispersion brings to the shift we calculate the ratio of the height of the dispersive peak to the nondispersive result at the same $\chi(0)$, and find
\begin{equation}
\frac{\Delta\mu_{\perp {\rm disp}}}{\Delta\mu_{\perp{\rm nondisp}}} \approx \frac{30.3 \,\text{eVnm}}{|\omega_T z|}\;, \quad 
\frac{\Delta\mu_{\parallel {\rm disp}}}{\Delta\mu_{\parallel {\rm nondisp}}} \approx \frac{81.6\,\text{eVnm}}{|\omega_T z|}\;. \label{DispShifts}
\end{equation}
A typical value for the frequency $\omega_T$ in a metal is on the order of a few eV (see, for example, \cite{SiliconDielectricFunction}), meaning that a significant enhancement relative to the nondispersive case would be observed only at extremely small distances $z$. However, restricting oneself to considering the properties of only elemental solids would be short-sighted. Structures engineered on the nanoscale can have transverse resonance frequencies $\omega_T$ significantly smaller than any ordinary material --- examples include an InSb semiconductor grating with $\omega_T$ (and $\omega_p$) in the range of a few meV \cite{SemiconductorPlasma}. These types of materials are at a focal point of strong contemporary interest in low-frequency plasmonics. With appropriate assumptions about the approximation of a part of such a structure as a planar surface, we find that for distances $z$ of a few tens of nanometres one may get an enhancement factor on the order of $10^3$ relative to the nondispersive case.

The apparent problem of the behaviour of the nondispersive result in the limit of large refractive index, $n\to\infty$, can be clarified by comparing it with the behaviour of the dispersive shift at large $\chi(0)$. In this regime the shift for the dispersive dielectric model becomes linear in $\sqrt{\chi(0)}$ and agrees with the nondispersive results; so for large $\chi(0)$ the two models are equivalent. The crucial additional observation is to note that for a nondispersive dielectric with large $\chi(0)$ we have $\chi(0)\approx n^2$, which is to say that a large the refractive index necessarily implies a large static susceptibility. Therefore, in the nondispersive model one cannot sensibly make a distinction between an arbitrarily large refractive index and an arbitrarily large static susceptibility. Investigation of the dispersive dielectric has shown that the latter interpretation is the correct one --- the magnetic moment shift grows with increasing static susceptibility, but an arbitrarily large static susceptibility is, of course, physically impossible. So while the shift in the nondispersive case does indeed increase without bound as the refractive index $n$ is increased, this is not due any problem with the calculation, but is in fact the result of the static susceptibility growing without bound
and an inevitable consequence of the unrealistic exclusion of dispersion from the model. 

Consideration of the shifts in terms of the static susceptibility also emphasizes the close relationship between plasma and perfect reflector models. In both of these models the static susceptibility is infinite right from the start, which means that their results do agree in the limit $\omega_p \to \infty$.

\begin{figure}
\includegraphics[width = 8cm]{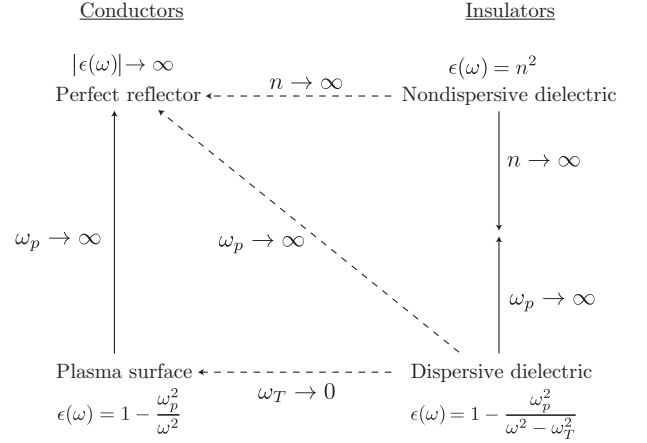}
\caption{\label{fig:commutation} Commutation properties of the various models discussed in the text. Each arrow indicates a limit which takes one dielectric function to another. Solid (dashed) arrows indicate a limit that, when applied the reflection coefficients, commutes (does not commute) with the limit $k_z \to -ik_\parallel$. The consequence of this is that the magnetic moment results for two models connected by solid (dashed) arrows are (are not) obtainable as limiting cases of one another. }
\end{figure}

The differences between the four models discussed above very clearly show that in order to predict the magnetic moment shift for a given set-up, one must choose a model which is physically appropriate for the low-frequency behaviour of electromagnetic response of the material at hand. In other words, it matters whether the material is a conductor or an insulator. These two classes of material are not obtainable as limiting cases of each other because the conductor models ignore the existence of evanescent modes (which is a direct consequence of their static susceptibilities being infinite). The calculations for each class of model diverge from each other because of non-commutation of a variety of limits of the reflection coefficient, namely between the static limit ($k_z \to -ik_\parallel$) and whichever limit we have to take in order to compare models. For example, we note that the $n\to\infty$ and $k_z \to -ik_\parallel$ limits of the nondispersive TE reflection coefficient do not commute, which leads to the $n\to\infty$ limit of the result for a nondispersive dielectric to disagree with the perfect reflector result. A further important example is that the limit of vanishing transverse resonance frequency, $\omega_T\to 0$, and the static limit $k_z \to -ik_\parallel$ of the dispersive TE reflection coefficient do not commute, which means that taking $\omega_T\to 0$ ($\chi(0) \to \infty$) in the dispersive dielectric results will not reproduce the plasma results, while naive comparison of the dielectric functions \eqref{PlasmaDielectricFunction} and \eqref{DispersiveDielectricFunction} suggests that they should. Since this plays such an important roles for the physical interpretation of the results, we summarize the commutation (or lack thereof) between the various limits of the reflection coefficients in Fig.~\ref{fig:commutation}.

For the plasma model, we found a $1/z^3$ dependence of the magnetic moment shift at small distances, i.e.~small $|\omega_p z|$, and that the leading $1/z^3$ term can be found either by determining the asymptotics of the complete shift, or by considering only the part due to the interaction with just surface plasmons. The asymptotics of the integrals for the shift in the dispersive dielectric case are too awkward to analyse directly. Instead we give the results one obtains by considering only the interaction with the surface polariton, i.e.~by using Eqs.~\eqref{DispersiveDielectricFunction} and \eqref{NewDispRelation} in Eqs.~\eqref{SPOnly_perp} and (\ref{SPOnly_para}):
\begin{align}
\Delta \mu_\perp (|\omega_p z| \ll1) &\approx \frac{e^3}{64 \pi \sqrt{2} m^3 z^3}\frac{1}{ \sqrt{2 \omega_T^2 + \omega_p^2}} \label{smallomegapzperp} 
\end{align}
\begin{align}
\Delta \mu_\parallel (|\omega_p z| \ll1) &\approx \frac{5e^3}{128 \pi \sqrt{2} m^3 z^3}\frac{1}{ \sqrt{2 \omega_T^2 + \omega_p^2}} \label{smallomegapzparallel}
\end{align} 
Surprisingly, for these short-distances $|\omega_p z|\ll 1$, we find that the $\omega_T\to 0$ limits of Eqs.~\eqref{smallomegapzperp} and \eqref{smallomegapzparallel} \emph{do} agree with the corresponding results for the plasma, Eqs.~\eqref{SmallzperpPlasma} and \eqref{SmallzparaPlasma}, unlike the results for general distances $|\omega_p z| \gtrsim 1$. This is because these results depends only on the surface plasmon part of the mode functions and electrostatic interactions, but there is no reflection of travelling photon modes, and hence any non-commutation of limits in the reflection coefficient does not come into play. 

\begin{figure}
\includegraphics[width=85mm]{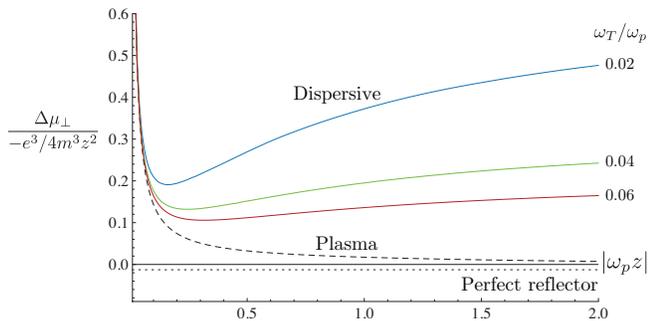}
\caption{\label{fig:zdep}(Color online)  Magnetic moment shift for dispersive dielectric and plasma models as a function of scaled distance $|\omega_pz|$ from the surface, for the case of the magnetic field ${\bf B}_0$ perpendicular to the surface and $\omega_T/\omega_p=\{0.02, 0.04, 0.06$.\}}
\end{figure}

At large distances $|\omega_p z|\gg 1$ the shift decreases as $1/z^2$, which is obvious when applying Watson's lemma to the integrals in Eqs.~(\ref{FullResultPerp}) and (\ref{FullResultPara}) or, more conveniently, to those in Eqs.~(\ref{OmegaResultPerp}) and (\ref{OmegaResultPara}). In fact, its asymptotics is given by the same expressions as in Eqs.~(\ref{DeltaMuPerp}) and (\ref{DeltaMuPara}) except with $n$ replaced by $\sqrt{1+(\omega_p/\omega_T)^2}$, i.e.~by the square root of the dielectric function at zero frequency, which is expected because at large distances only the static electromagnetic response of the surface matters and dispersion plays no role to leading order.
Figure \ref{fig:zdep} shows the shift for dispersive dielectric and plasma models as a function of distance.

\section{Experimental Relevance}\label{ExpRel}

Expressing magnetic moment shifts as relative shifts $\Delta \mu/\mu$ to the Dirac magnetic moment $\mu=e/2m$, we have for the perpendicular component of the nondispersive shift in SI units:
\begin{equation}
\frac{\Delta \mu_{\perp\text{nondisp}}}{\mu}=\frac{\hbar}{c^3 \epsilon_0} \frac{e^2}{16\pi^2 m^2 z^2} f(n) \approx \frac{10^{-11} \text{nm}^2}{z^2}
\label{relativeshift}
\end{equation}
where $f(n)$ is the remaining part of Eq.~\eqref{DeltaMuPerp}, and is of order unity. For a distance $z\approx\;$1nm,  Eqs.~\eqref{DispShifts} (and the discussions following them) show that the enhancement due to the inclusion of dispersion can be of order $10^4$ under favourable conditions. Thus, we have a magnetic moment shift of up to one part in $10^7$. The current experimental accuracy for $g/2$ in free space is on the order of one part in $10^{12}$ \cite{Odom}, so that the shift calculated here would compare very favourably to this. As the distance increases to the order of a micron the effect decreases towards the limits of current experimental accuracy. For example, an electron 0.1$\mu$m away from the same surface as above would have its magnetic moment shifted by only one part in $10^{11}$.

This leads one to ask if the current best techniques for measuring the g factor would be suitable for making a measurement of the surface dependent shift of the magnetic moment. Since one of the sticking points in such experiments is that accurate measurement of the externally applied magnetic field $B_0$ is mostly impossible, g-factor experiments usually do not measure the magnetic moment directly, but instead they find its ratio to either a known magnetic moment, or to the cyclotron frequency of the particle under consideration. In case of the latter for surface-dependent magnetic moments shifts one would need to take into account the shift in cyclotron frequency of a particle near a surface, which arises due to the position-dependent self-energy of the particle \cite{BartonFawcett, EberleinRobaschik, massshift}. Crucially, the leading term of the surface-dependent cyclotron frequency shift is of order $\alpha/(mz)$ and thus much bigger than the magnetic moment shift which is of order $\alpha/(mz)^2$ (cf.~Eq.~(\ref{relativeshift})). So, an experiment which adapts the techniques used for measuring the free-space g factor to find its surface dependant part would effectively be measuring the change due to the surface in its self-energy, not in its magnetic moment. While direct experimental confirmation of a shift in the self-energy would, of course, be interesting in its own right, its existence represents a significant obstacle to isolation and observation of the magnetic moment shift.

\section{Summary and conclusions}

We have shown that the magnetic moment shift near a nondispersive imperfectly reflecting surface has notable differences from the corresponding shift near a perfectly reflecting surface, differing by orders of magnitude and in some cases even in its sign. The inclusion of dispersion can significantly modify the magnitude of the effect, and that this modification can be tuned by choice of material. We have given a general formula for the calculation of magnetic moment shifts, into which one can simply insert the relevant dielectric function and evaluate the integrals to obtain the shift as dependent on the distance from the surface and various parameters characterizing the electromagnetic response of the surface. 

The work presented here has also considerably extended the horizon of the traditional theory of cavity quantum electrodynamics.
While the effect that causes the shift of the magnetic moment of an electron close to a surface is essentially the same as the one that causes the Casimir-Polder shift of energy levels in neutral atoms close to a surface, the theory required for the magnetic moment shift is considerably more involved. On the one hand, this is because the nature of the spin requires a relativistic treatment, but on the other hand, the much more subtle but ultimately more important reason is that dynamics of the system is qualitatively more complicated: a neutral atom interacting with a surface involves the exchange of virtual excitations between just the photon field and the atomic electron, whereas the shift of magnetic moment of an electron near a surface involves virtual excitations being exchanged in a triangle between the photon field, the centre-of-mass motion of the electron, and its spin states. This is also the reason as to why there is no simple resonance effect when the spin-flip frequency in a magnetic field coincides with an absorption resonance in the material of the surface.

\begin{acknowledgments}
It is a pleasure to thank Robert Zietal for discussions. Financial support from the UK Engineering and Physical Sciences Council (EPSRC) is gratefully acknowledged.
\end{acknowledgments}

\appendix
\section{Schr\"odinger and Dirac equations for a particle in a constant magnetic field}\label{LandauQ}
In order to derive the eigenstates of the Dirac equation for a particle in a constant classical magnetic field  $\B{B}_0$ one first solves the corresponding Schr\"{o}dinger problem and then uses its solutions to generate the Dirac eigenstates \cite{JohnsonLippmann}. 
The Schr\"{o}dinger Hamiltonian for a charged particle moving in a constant magnetic field $\B{B}_0=B_0 \hat{z}$ is
\begin{equation}
H_S = \frac{(p_x+\frac{eB_0}{2}y)^2}{2m}+ \frac{(p_y-\frac{eB_0}{2}x)^2}{2m}+\frac{p_z^2}{2m}
\end{equation}
One can reduce this to a harmonic oscillator by introducing annihilation and creation operators and rewriting the positions and momenta in terms of those,
\begin{align}
\hat{x} &=\frac{1}{\beta_0 \sqrt{2}}(\bx+\bxd) &\hat{p}_x &= \frac{i\beta_0}{\sqrt{2}}(\bxd-\bx)\\
\hat{y}&=\frac{1}{\beta_0 \sqrt{2}}(\by+\byd)& \hat{p}_y&=\frac{i\beta_0}{\sqrt{2}}(\byd-\by)
\end{align}
where $\beta_0=\sqrt{-eB_0/2}$. (Note that we use $e=-|e|$). The operators $\bx$ and $\by$ are then combined to form creation and annihilation operators for right and left-circular quanta
\begin{align*}
\bR&=\frac{1}{\sqrt{2}} (\bx-i\by)\;,\ \ \bRd=\frac{1}{\sqrt{2}} (\bxd+i\byd)\;,\\
\bL&=\frac{1}{\sqrt{2}} (\bx+i\by)\;,\ \ \bLd=\frac{1}{\sqrt{2}} (\bxd-i\byd)\;.
\end{align*}
In terms of these the canonical momenta are then given by
\begin{align*}
\hat{\pi}_x &= \hat{p}_x+\frac{eB_0}{2}\hat{y} = i\beta_0 (\bRd-\bR)\\
\hat{\pi}_y &= \hat{p}_y-\frac{eB_0}{2}\hat{x} = \beta_0 (\bRd+\bR)\\
\end{align*}
so that the Hamiltonian reads as
 \begin{equation}
H_S=-\frac{eB_0}{m}\left(\bRd \bR+\frac{1}{2}\right)+\frac{p_z^2}{2m}\;.
\label{HS}
\end{equation}
Thus the Hamiltonian is equivalent to a harmonic oscillator of right-circular excitations and possesses infinite degeneracy with respect to the left-circular quanta. Eigenstates $|\nu\rangle$ of the Schr\"{o}dinger Hamiltonian $H_S$ can therefore be generated by repeated application of the creation operator $\bRd$ to the ground state $|\nu=0\rangle$ which is defined by $\bR|\nu=0\rangle =0$.

We can now use the Schr\"{o}dinger eigenstates to derive the corresponding Dirac eigenstates. Following \cite{JohnsonLippmann}, we start by noting that eigenfunctions of the Dirac equation
\begin{equation}
(\boldsymbol{\alpha}\cdot\boldsymbol{\pi}+\beta m)\psi \equiv H_0\psi = E_\nu \psi \label{DiracEqnE}
\end{equation}
may be obtained from solutions of
\begin{equation}
(H_0^2-E_\nu^2)X=(H_0-E_\nu)(H_0+E_\nu) X =0\;. \label{DiracEqnX}
\end{equation}
Evidently, if a state $X$ satisfies the above equation, then 
\begin{equation}
\psi = (H_0+E_\nu)X 
\end{equation}
is a solution of the Dirac eigenvalue problem, Eq.~(\ref{DiracEqnE}). To find the eigenvalues $E_\nu ^2$ of $H_0^2$ we calculate $H_0^2$, which on account of $(\boldsymbol{\alpha}\cdot\boldsymbol{\pi})^2=\boldsymbol{\pi}^2 - e\sigma_z B_0 $ and the anticommutation properties of the Dirac matrices, reads as
\begin{equation}
H_0^2= \boldsymbol{\pi}^2 - e\sigma_z B_0 +m^2\;.
\end{equation}
Therefore we can relate $H_0^2$ to the Schr\"odinger Hamiltonian $H_S$,
\begin{equation}
H_0^2= 2mH_S - e\sigma_z B_0 +m^2\;,
\label{Dirac_from_Schroed}
\end{equation}
and then the eigenvalues $E_\nu ^2$ of $H_0^2$ follow from Eq.~(\ref{HS}) and the eigenvalues $s$ of the spin operator $S_z=\sigma_z/2$,
\begin{equation}
E_\nu ^2=m^2+p_z^2 - 2eB_0\left(\nu +s+\frac{1}{2}\right)
\label{E_squared}
\end{equation}
We now choose the states $X$ in such a way that they distinguish spin-up and spin-down states, and particle and anti-particle states, i.e. we choose them to be eigenfunctions of $\sigma_z$ with eigenvalues $s=\pm 1/2$, and of $\beta\equiv\gamma_0$ with eigenvalues $1$ for a particle and $-1$ for an antiparticle. 
Equation (\ref{Dirac_from_Schroed}) implies that the Dirac eigenstates can be expressed in terms of a product state of the non-relativistic eigenstates $\ket{\nu}$ and the spin state $\ket{s}$, which we choose to write as $\ket{\nu,s}\equiv\ket{\nu}\otimes\ket{s}$,
\begin{equation}
\ket{\Psi_e } = \frac{H_0+E_\nu }{\sqrt{2E_\nu (E_\nu +m)}}\ket{\nu } \chi^{(\uparrow,\downarrow)} \ \ \text{for $s=\pm\nicefrac{1}{2}$,}
\label{eigenstate}
\end{equation}
where $\chi^{(\uparrow)\dagger} = (1, 0, 0, 0)$, $\chi^{(\downarrow)\dagger} = (0, 1, 0, 0)$. For antiparticle eigenstates the negative root of (\ref{E_squared}) applies, the normalization factor in the denominator of Eq.~(\ref{eigenstate}) turns into $\sqrt{-2E_\nu (-E_\nu +m)}$, 
 and we use $\chi^{(\uparrow)\dagger} = (0, 0, 1, 0)$, $\chi^{(\downarrow)\dagger} = (0, 0, 0, 1)$. 

For calculations it is expedient to express momentum components in terms of
\begin{align}
\pi_{+} &= \pi_x + i\pi_y = 2i\beta_0 \hat{b}_R^\dagger\;, \\
\pi_{-}  &= \pi_x - i\pi_y =-2i\beta_0 \hat{b}_R\;.
\end{align}
Thus, for a general vector $\mathbf{Q}$ we have
\begin{equation}
\mathbf{Q}\cdot \boldsymbol{\pi} = i \beta_0 Q_{-} \hat{b}_R^\dagger -  i \beta_0 Q_{+} \hat{b}_R + Q_z p_z \label{QOperator}
\end{equation}
with 
\begin{align*}
 Q_{+}&=Q_x+iQ_y\\
 Q_{-}&=Q_x-iQ_y
\end{align*}

In the main text we also need the matrix elements of the position operator. Noting that \cite{CohenTannoudji}
\begin{align}
x=\frac{1}{2\beta_0}(\hat{b}_R+\hat{b}_L+\bRd+\bLd)=x_0+\frac{1}{2\beta_0}(\hat{b}_R+\bRd)\notag\\
y=\frac{i}{2\beta_0}(\hat{b}_R-\hat{b}_L-\bRd+\bLd)=y_0+\frac{i}{2\beta_0}(\hat{b}_R-\bRd)\;, \notag
\end{align}
we have
\begin{eqnarray}
\bra{\nu +1}(\hat{x}-\hat{x}_0)\ket{\nu } &=& \frac{1}{2\beta_0}\sqrt{\nu +1} \nonumber \\
\bra{\nu -1}(\hat{x}-\hat{x}_0)\ket{\nu } &=& \frac{1}{2\beta_0}\sqrt{\nu } \nonumber \\
\bra{\nu +1}(\hat{y}-\hat{y}_0)\ket{\nu } &=& -\frac{i}{2\beta_0}\sqrt{\nu +1} \nonumber \\
\bra{\nu -1}(\hat{y}-\hat{y}_0)\ket{\nu } &=& \frac{i}{2\beta_0}\sqrt{\nu }\;.
\label{position_elements}
\end{eqnarray}

\vspace*{2mm}
\section{Reproduction of perfect mirror result} \label{PerfectMirrorCalc}

In the derivation of Eqs.~(\ref{FullResultPerp})and (\ref{FullResultPara}) it was argued that the subtraction and re-addition of the point $\{\xi\to 0 ,\eta \to \infty\}$ from the TE reflection coefficients would not be necessary since $R^L_{\text{TE}}$ is zero at this point for any physically reasonable dielectric function. However, this does not apply to the perfect reflector which has a TE reflection coefficient of $-1$ at all frequencies, so that Eqs.~(\ref{FullResultPerp}) and (\ref{FullResultPara}) cannot be used for calculating the magnetic moment shift near a perfect reflector. Instead, we go one step back and start from Eqs.~(\ref{ShiftAfterContourPerp}) and (\ref{ShiftAfterContourPara}).
At first glance it appears that the branch cut due to $k_z^d$ might meet that due to $k$ when the limit $n \to \infty$ is taken. However, in fact the branch cut due to $k_z^d$ disappears since the perfect reflector excludes all right-incident modes, so that $k_z^d$ does not appear in the integrand.  Therefore the contour of integration simply runs straight along the $k_z$ axis. This can also be seen formally by using the integral over $C'$, taking the limit $n\to\infty$ (which eliminates the branch cut) and then deforming the contour back up to the real axis. Either way one obtains
\begin{align}
\Delta \mu_{\perp,PM} &=-\frac{e^3}{32\pi^2m^3}  \int_0^\infty dk_\parallel \int_{-\infty}^\infty dk_z \frac{k_\parallel}{k^3} \nonumber\\&\times\left[ \left( 2k_\parallel^2-k_z^2  \right)(-1)+\left(2k_\parallel^2+k_z^2 \right)(+1)\right] e^{2ik_zz} \nonumber\\&= \frac{e^3}{32 \pi^2 m^3 z^2 } \;,\label{PM1}\\
\Delta \mu_{\parallel,PM} &=-\frac{e^3}{32\pi^2m^3}  \int_0^\infty dk_\parallel \int_{-\infty}^\infty dk_z \frac{k_\parallel}{2k^3} \nonumber\\&\times\left[ \left( 3k_\parallel^2+2k_z^2  \right)(-1)+\left(3k_\parallel^2-2k_z^2 \right)(+1)\right] e^{2ik_zz} \nonumber\\&= -\frac{e^3}{32 \pi^2 m^3 z^2 } \;,\label{PM2}
\end{align}
where we have carried out the $k_z$ integration first and used
\begin{equation}
\int_{-\infty}^\infty dk_z \frac{k_\parallel\;e^{2ik_zz}}{\left(k_z^2+k_\parallel^2\right)^{3/2}} = 4|z|\; K_1(2k_\parallel |z|)\;.
\end{equation}
Eqs.~(\ref{PM1}) and (\ref{PM2}) reproduce the shifts calculated by \cite{BartonFawcett}, and the $n$-independent terms in Eqs.~(\ref{Largennondisp}) and (\ref{LargennondispRotated}). This calculation shows another way of looking at the fundamental disparity between the perfect reflector and nondispersive dielectric models: the branch cut due to $k_z^d$ meets that due to $k$ if $n$ is taken to be finite at the start of the calculation and then made infinite at the end, while if $n$ is infinite from the start the branch cut never appears in the first place.

\section{TE Part of Plasma Surface} \label{TEPlasmaAppendix} 
Using Eq.~(\ref{ShiftAfterContourPerp}) but with the integration contour $C'$ changed back to $C$ for the plasma model \emph{before} deformation (cf.~Fig.~\ref{fig:contour}), one can write the TE part of the shift for the plasma surface as
\begin{widetext}
\begin{equation}
\Delta \mu_{\perp,TE}= -\frac{e^3}{32\pi^2\omega_p^2m^3 } \int_0^\infty dk_\parallel \int_{-\infty}^\infty dk_z \, \frac{k_\parallel(2k_\parallel^2-k_z^2)}{(k_\parallel^2+k_z^2)^{3/2}} \left(2k_z^2- \omega_p^2 - 2k_z\sqrt{k_z^2 - \omega_p^2}\right)e^{2ik_z z}
\end{equation}
where the reflection coefficient has been written out explicitly using Eq.~(\ref{Plasmakzd}) and the branch cut is taken to be between the branch points at $\pm\omega_p$. Care must be taken when evaluating this integral due to the physical requirement $\text{sgn}(k_z) = \text{sgn}(k_z^d)$ from refraction; so we outline the calculation here. First, we note that the order of integration matters, and that the integral is convergent only if the $k_z$ integration is carried out first. To circumvent this problem we introduce a cutoff $\Lambda$ on the $k_\parallel$ integral. This improves the convergence of the double integral so that we are allowed to interchange the order of integrations. The $k_\parallel$ integral can then be calculated exactly and gives:
\begin{equation}
\Delta \mu_{\perp,TE}=-\frac{e^3}{32\pi^2\omega_p^2m^3 }\lim_{\Lambda\to\infty}\left[ \int_{-\infty}^\infty dk_z \, (2\Lambda-5|k_z|) \left(2k_z^2- \omega_p^2 - 2k_z\sqrt{k_z^2 - \omega_p^2}\right)e^{2ik_z z} + \mathcal{O}(1/\Lambda)\right] \label{TEPlasmaStart}.
\end{equation}
We first consider the term with the square root. For the contribution from the region $|k_z|>\omega_p$, we have (omitting the overall constants):
\begin{equation}
\Bigg\{ \int_{-\infty}^{-\omega_p}  dk_z + \int_{\omega_p}^{\infty} dk_z \Bigg\}\; e^{2ik_zz} 
 (2\Lambda-5|k_z|)\left( - 2k_z\sqrt{k_z^2 - \omega_p^2}\right) 
\end{equation}
\end{widetext}
Noting that $k_z\sqrt{k_z^2 - \omega_p^2}$ is even in $k_z$ because of the physical constraint $\text{sgn}(k_z)=\text{sgn}\sqrt{k_z^2 - \omega_p^2}$, we can simplify this to:
\begin{equation}
=2\int_{\omega_p}^{\infty} dk_z \cos(2k_z z) (2\Lambda-5|k_z|)\left( - 2k_z\sqrt{k_z^2 - \omega_p^2}\right)\;. \label{TEOutside}
\end{equation}
Next we consider the region $|k_z|<\omega_p$, where $k_z^d=\sqrt{k_z^2 - \omega_p^2}$ is imaginary. As illustrated in Fig.~\ref{fig:contour}, the integration path runs along one and the same side of the cut, which means that the factor $k_z\sqrt{\omega_p^2-k_z^2 }$ is now odd in $k_z$. 
Applying the constraint $\text{sgn}(k_z)=\text{sgn}\sqrt{k_z^2 - \omega_p^2}$ to the vicinity of $k_z\approx \omega_p$, we are directed to choosing the complex sheet of the square root such that in the lower half-plane $\sqrt{k_z^2 - \omega_p^2} = -i\sqrt{ \omega_p^2-k_z^2}$. These considerations lead us to write the integral analogous to Eq.~(\ref{TEOutside}) but from the region $|k_z|<\omega_p$ as
\begin{equation}
2\int_{0}^{\omega_p} dk_z \, (2\Lambda-5k_z)\left( - 2k_z\sqrt{\omega_p^2-k_z^2 }\right) \sin(2k_zz) \label{TEInside}
\end{equation}
The rest of Eq.~(\ref{TEPlasmaStart}) is a trivial integral, and combining this with Eqs.~(\ref{TEOutside}) and (\ref{TEInside}) gives:
\begin{widetext}
\begin{align}
=  -\frac{e^3}{16\pi^2m^3 }\lim_{\Lambda \to \infty} \Bigg\{ \frac{1}{\omega_p^2} \int_{0}^{\omega_p} dk_z \,&\left[(2k_z^2 - \omega_p^2)\cos(2k_zz) - 2k_z\sqrt{\omega_p^2-k_z^2 } \sin(2k_zz) \right] (2\Lambda -5k_z)\notag  \\ 
+ &\frac{1}{\omega_p^2}\int_{\omega_p}^{\infty} dk_z  \cos(2k_z z)\left(2k_z^2 - \omega_p^2 - 2k_z\sqrt{k_z^2 - \omega_p^2}\right) (2\Lambda -5k_z)\Bigg\}.
\end{align}
The integrals proportional to $\Lambda$ give expressions with the Bessel function $J_2(2\omega_pz)$, $\sin(2\omega_pz)$, and $\cos(2\omega_pz)$, but all together they conspire to add up to zero. Defining 
\begin{align}
\mathcal{I}_{TE}\equiv \frac{1}{{\omega_p^2}} \Bigg\{ \int_{0}^{\omega_p}\! dk_z  k_z \Big[(2k_z^2 - \omega_p^2)&\cos(2k_zz) 
- 2k_z\sqrt{\omega_p^2-k_z^2 } \sin(2k_zz) \Big] \notag \\
&+ \int_{\omega_p}^{\infty} dk_z \, k_z \cos(2k_z z)\left((2k_z^2 - \omega_p^2) - 2k_z\sqrt{k_z^2 - \omega_p^2}\right) \Bigg\}
\end{align}
\end{widetext}
we therefore have
\begin{equation}
\Delta \mu_{\perp,TE}=\frac{5e^3}{16\pi^2m^3 }\mathcal{I}_{TE} \quad \mbox{and} \quad \Delta \mu_{\parallel,TE}=\frac{e^3}{8\pi^2m^3 }\mathcal{I}_{TE}\;,
\end{equation}
where the case for ${\bf B}_0$ parallel to the interface has been evaluated in exactly the same way. The integral $\mathcal{I}_{TE}$ may be evaluated analytically; one finds
\begin{align}
\mathcal{I}_{TE}&=\frac{1}{4 z^2}+\frac{3}{4 z^4 \omega_p ^2}-\frac{4 z \omega_p ^3}{15}+\frac{\pi  \omega_p  Y_1(-2 \omega_p z)}{2 z}\notag \\&+\frac{3 \pi  Y_2(-2 \omega_p z)}{4 z^2}-\frac{\pi  H_2(2  \omega_p z )}{4 z^2}+\frac{\pi  \omega_p  H_3(2  \omega_p z )}{2 z}
\end{align}
where $Y_n$ is the $n$th Bessel function of the second kind, and $H_n$ is the $n$th Struve function. This result displays the expected behaviour that $\lim_{z\to -\infty}\mathcal{I}_{TE} = 0$, i.e.~ that there is no magnetic moment shift due to a surface that is infinitely far away. 

The `perfect-mirror' limit of this object is
\begin{equation}
 \lim_{\omega_p\to \infty}\mathcal{I}_{TE} = \frac{1}{4z^2}
\end{equation}
which means that for the plasma surface the TE modes do not result in unlimited growth of the magnetic moment shift for large values of the permittivity, in contrast to what was observed for the nondispersive dielectric in Eqs.~(\ref{Largennondisp}) and (\ref{LargennondispRotated}).

Inspection of Eqs.~(\ref{plasmaTMPerp}) and (\ref{plasmaTMPara}) in the limit $|\omega_pz|\to\infty$ shows that the $s$ integrals give rise to terms proportional to $1/(\omega_pz^3)$, whence only the $1/z^2$ terms in front of them contribute to leading order. We then have in total
\begin{align*}
\Delta \mu_\perp(\omega_p \to \infty) &= \frac{e^3}{16 \pi^2 m^3}\left( \frac{5}{4z^2}-\frac{3}{4z^2}\right) = \frac{e^3}{32 \pi^2 m^3z^2}\\
\Delta \mu_\parallel(\omega_p \to \infty) &= \frac{e^3}{16 \pi^2 m^3}\left( \frac{1}{2z^2}-\frac{1}{z^2}\right) = -\frac{e^3}{32 \pi^2 m^3z^2}
\end{align*} 
in agreement with the perfect-mirror limit, and also with the $n$-independent terms from Eqs.~(\ref{Largennondisp}) and (\ref{LargennondispRotated}).

We note that the asymptotics of $I_{TE}$ for small $|\omega_p z|$ are
\begin{equation}
\mathcal{I}_{TE}(|\omega_p z| \ll 1) =  -\frac{\omega_p^2}{16}  \left[1+4 \gamma +4 \ln(-\omega_p z)\right] \label{ITESmallwpz}
\end{equation}
where $\gamma$ is the Euler constant $\approx 0.577$. 


\begin{thebibliography}{31}%
\makeatletter
\providecommand \@ifxundefined [1]{%
 \@ifx{#1\undefined}
}%
\providecommand \@ifnum [1]{%
 \ifnum #1\expandafter \@firstoftwo
 \else \expandafter \@secondoftwo
 \fi
}%
\providecommand \@ifx [1]{%
 \ifx #1\expandafter \@firstoftwo
 \else \expandafter \@secondoftwo
 \fi
}%
\providecommand \natexlab [1]{#1}%
\providecommand \enquote  [1]{``#1''}%
\providecommand \bibnamefont  [1]{#1}%
\providecommand \bibfnamefont [1]{#1}%
\providecommand \citenamefont [1]{#1}%
\providecommand \href@noop [0]{\@secondoftwo}%
\providecommand \href [0]{\begingroup \@sanitize@url \@href}%
\providecommand \@href[1]{\@@startlink{#1}\@@href}%
\providecommand \@@href[1]{\endgroup#1\@@endlink}%
\providecommand \@sanitize@url [0]{\catcode `\\12\catcode `\$12\catcode
  `\&12\catcode `\#12\catcode `\^12\catcode `\_12\catcode `\%12\relax}%
\providecommand \@@startlink[1]{}%
\providecommand \@@endlink[0]{}%
\providecommand \url  [0]{\begingroup\@sanitize@url \@url }%
\providecommand \@url [1]{\endgroup\@href {#1}{\urlprefix }}%
\providecommand \urlprefix  [0]{URL }%
\providecommand \Eprint [0]{\href }%
\providecommand \doibase [0]{http://dx.doi.org/}%
\providecommand \selectlanguage [0]{\@gobble}%
\providecommand \bibinfo  [0]{\@secondoftwo}%
\providecommand \bibfield  [0]{\@secondoftwo}%
\providecommand \translation [1]{[#1]}%
\providecommand \BibitemOpen [0]{}%
\providecommand \bibitemStop [0]{}%
\providecommand \bibitemNoStop [0]{.\EOS\space}%
\providecommand \EOS [0]{\spacefactor3000\relax}%
\providecommand \BibitemShut  [1]{\csname bibitem#1\endcsname}%
\let\auto@bib@innerbib\@empty
\bibitem [{\citenamefont {Odom}\ \emph {et~al.}(2006)\citenamefont {Odom},
  \citenamefont {Hanneke}, \citenamefont {D'Urso},\ and\ \citenamefont
  {Gabrielse}}]{Odom}%
  \BibitemOpen
  \bibfield  {author} {\bibinfo {author} {\bibfnamefont {B.}~\bibnamefont
  {Odom}}, \bibinfo {author} {\bibfnamefont {D.}~\bibnamefont {Hanneke}},
  \bibinfo {author} {\bibfnamefont {B.}~\bibnamefont {D'Urso}}, \ and\ \bibinfo
  {author} {\bibfnamefont {G.}~\bibnamefont {Gabrielse}},\ }\href {\doibase
  10.1103/PhysRevLett.97.030801} {\bibfield  {journal} {\bibinfo  {journal}
  {Phys. Rev. Lett.}\ }\textbf {\bibinfo {volume} {97}},\ \bibinfo {pages}
  {030801} (\bibinfo {year} {2006})}\BibitemShut {NoStop}%
\bibitem [{\citenamefont {Hanneke}\ \emph {et~al.}(2008)\citenamefont
  {Hanneke}, \citenamefont {Fogwell},\ and\ \citenamefont
  {Gabrielse}}]{Hanneke}%
  \BibitemOpen
  \bibfield  {author} {\bibinfo {author} {\bibfnamefont {D.}~\bibnamefont
  {Hanneke}}, \bibinfo {author} {\bibfnamefont {S.}~\bibnamefont {Fogwell}}, \
  and\ \bibinfo {author} {\bibfnamefont {G.}~\bibnamefont {Gabrielse}},\ }\href
  {\doibase 10.1103/PhysRevLett.100.120801} {\bibfield  {journal} {\bibinfo
  {journal} {Phys. Rev. Lett.}\ }\textbf {\bibinfo {volume} {100}},\ \bibinfo
  {pages} {120801} (\bibinfo {year} {2008})}\BibitemShut {NoStop}%
\bibitem [{\citenamefont {Fischbach}\ and\ \citenamefont
  {Nakagawa}(1984)}]{Fischbach}%
  \BibitemOpen
  \bibfield  {author} {\bibinfo {author} {\bibfnamefont {E.}~\bibnamefont
  {Fischbach}}\ and\ \bibinfo {author} {\bibfnamefont {N.}~\bibnamefont
  {Nakagawa}},\ }\href {\doibase 10.1103/PhysRevD.30.2356} {\bibfield
  {journal} {\bibinfo  {journal} {Phys. Rev. D}\ }\textbf {\bibinfo {volume}
  {30}},\ \bibinfo {pages} {2356} (\bibinfo {year} {1984})}\BibitemShut
  {NoStop}%
\bibitem [{\citenamefont {Svozil}(1985)}]{Svozil}%
  \BibitemOpen
  \bibfield  {author} {\bibinfo {author} {\bibfnamefont {K.}~\bibnamefont
  {Svozil}},\ }\href {\doibase 10.1103/PhysRevLett.54.742} {\bibfield
  {journal} {\bibinfo  {journal} {Phys. Rev. Lett.}\ }\textbf {\bibinfo
  {volume} {54}},\ \bibinfo {pages} {742} (\bibinfo {year} {1985})}\BibitemShut
  {NoStop}%
\bibitem [{\citenamefont {Bordag}(1986)}]{Bordag}%
  \BibitemOpen
  \bibfield  {author} {\bibinfo {author} {\bibfnamefont {M.}~\bibnamefont
  {Bordag}},\ }\href {\doibase 10.1016/0370-2693(86)91009-9} {\bibfield
  {journal} {\bibinfo  {journal} {Phys. Lett. B}\ }\textbf {\bibinfo {volume}
  {171}},\ \bibinfo {pages} {113 } (\bibinfo {year} {1986})}\BibitemShut
  {NoStop}%
\bibitem [{\citenamefont {Boulware}\ \emph {et~al.}(1985)\citenamefont
  {Boulware}, \citenamefont {Brown},\ and\ \citenamefont
  {Lee}}]{BoulwareBrown}%
  \BibitemOpen
  \bibfield  {author} {\bibinfo {author} {\bibfnamefont {D.~G.}\ \bibnamefont
  {Boulware}}, \bibinfo {author} {\bibfnamefont {L.~S.}\ \bibnamefont {Brown}},
  \ and\ \bibinfo {author} {\bibfnamefont {T.}~\bibnamefont {Lee}},\ }\href
  {\doibase 10.1103/PhysRevD.32.729} {\bibfield  {journal} {\bibinfo  {journal}
  {Phys. Rev. D}\ }\textbf {\bibinfo {volume} {32}},\ \bibinfo {pages} {729}
  (\bibinfo {year} {1985})}\BibitemShut {NoStop}%
\bibitem [{\citenamefont {Kreuzer}\ and\ \citenamefont
  {Svozil}(1986)}]{KreuzerSvozil}%
  \BibitemOpen
  \bibfield  {author} {\bibinfo {author} {\bibfnamefont {M.}~\bibnamefont
  {Kreuzer}}\ and\ \bibinfo {author} {\bibfnamefont {K.}~\bibnamefont
  {Svozil}},\ }\href {\doibase 10.1103/PhysRevD.34.1429} {\bibfield  {journal}
  {\bibinfo  {journal} {Phys. Rev. D}\ }\textbf {\bibinfo {volume} {34}},\
  \bibinfo {pages} {1429} (\bibinfo {year} {1986})}\BibitemShut {NoStop}%
\bibitem [{\citenamefont {Kreuzer}(1988)}]{Kreuzer}%
  \BibitemOpen
  \bibfield  {author} {\bibinfo {author} {\bibfnamefont {M.}~\bibnamefont
  {Kreuzer}},\ }\href {\doibase 10.1088/0305-4470/21/15/017} {\bibfield
  {journal} {\bibinfo  {journal} {J. Phys. A: Math. Gen.}\ }\textbf {\bibinfo
  {volume} {21}},\ \bibinfo {pages} {3285} (\bibinfo {year}
  {1988})}\BibitemShut {NoStop}%
\bibitem [{\citenamefont {Barton}\ and\ \citenamefont
  {Fawcett}(1988)}]{BartonFawcett}%
  \BibitemOpen
  \bibfield  {author} {\bibinfo {author} {\bibfnamefont {G.}~\bibnamefont
  {Barton}}\ and\ \bibinfo {author} {\bibfnamefont {N.}~\bibnamefont
  {Fawcett}},\ }\href@noop {} {\bibfield  {journal} {\bibinfo  {journal}
  {Physics Reports}\ }\textbf {\bibinfo {volume} {170}},\ \bibinfo {pages} {1}
  (\bibinfo {year} {1988})}\BibitemShut {NoStop}%
\bibitem [{\citenamefont {Casimir}\ and\ \citenamefont {Polder}(1948)}]{CP}%
  \BibitemOpen
  \bibfield  {author} {\bibinfo {author} {\bibfnamefont {H.~B.~G.}\
  \bibnamefont {Casimir}}\ and\ \bibinfo {author} {\bibfnamefont
  {D.}~\bibnamefont {Polder}},\ }\href {\doibase 10.1103/PhysRev.73.360}
  {\bibfield  {journal} {\bibinfo  {journal} {Phys. Rev.}\ }\textbf {\bibinfo
  {volume} {73}},\ \bibinfo {pages} {360} (\bibinfo {year} {1948})}\BibitemShut
  {NoStop}%
\bibitem [{\citenamefont {Wu}\ and\ \citenamefont {Eberlein}(1999)}]{Wu}%
  \BibitemOpen
  \bibfield  {author} {\bibinfo {author} {\bibfnamefont {S.~T.}\ \bibnamefont
  {Wu}}\ and\ \bibinfo {author} {\bibfnamefont {C.}~\bibnamefont {Eberlein}},\
  }\href@noop {} {\bibfield  {journal} {\bibinfo  {journal} {Proc. R. Soc.
  Lond. A}\ }\textbf {\bibinfo {volume} {455}},\ \bibinfo {pages} {2487}
  (\bibinfo {year} {1999})}\BibitemShut {NoStop}%
\bibitem [{\citenamefont {Babiker}\ and\ \citenamefont
  {Barton}(1976)}]{BabikerBarton}%
  \BibitemOpen
  \bibfield  {author} {\bibinfo {author} {\bibfnamefont {M.}~\bibnamefont
  {Babiker}}\ and\ \bibinfo {author} {\bibfnamefont {G.}~\bibnamefont
  {Barton}},\ }\href@noop {} {\bibfield  {journal} {\bibinfo  {journal} {J.
  Phys. A: Math. Gen.}\ }\textbf {\bibinfo {volume} {9}},\ \bibinfo {pages}
  {129} (\bibinfo {year} {1976})}\BibitemShut {NoStop}%
\bibitem [{\citenamefont {Eberlein}\ and\ \citenamefont
  {Robaschik}(2004)}]{EberleinRobaschik}%
  \BibitemOpen
  \bibfield  {author} {\bibinfo {author} {\bibfnamefont {C.}~\bibnamefont
  {Eberlein}}\ and\ \bibinfo {author} {\bibfnamefont {D.}~\bibnamefont
  {Robaschik}},\ }\href {\doibase 10.1103/PhysRevLett.92.233602} {\bibfield
  {journal} {\bibinfo  {journal} {Phys. Rev. Lett.}\ }\textbf {\bibinfo
  {volume} {92}},\ \bibinfo {pages} {233602} (\bibinfo {year}
  {2004})}\BibitemShut {NoStop}%
\bibitem [{\citenamefont {Bennett}\ and\ \citenamefont
  {Eberlein}(2012)}]{massshift}%
  \BibitemOpen
  \bibfield  {author} {\bibinfo {author} {\bibfnamefont {R.}~\bibnamefont
  {Bennett}}\ and\ \bibinfo {author} {\bibfnamefont {C.}~\bibnamefont
  {Eberlein}},\ }\href {\doibase 10.1103/PhysRevA.86.062505} {\bibfield
  {journal} {\bibinfo  {journal} {Phys. Rev. A}\ }\textbf {\bibinfo {volume}
  {86}},\ \bibinfo {pages} {062505} (\bibinfo {year} {2012})}\BibitemShut
  {NoStop}%
\bibitem [{\citenamefont {Peskin}\ and\ \citenamefont
  {Schroeder}(1995)}]{Peskin}%
  \BibitemOpen
  \bibfield  {author} {\bibinfo {author} {\bibfnamefont {M.~E.}\ \bibnamefont
  {Peskin}}\ and\ \bibinfo {author} {\bibfnamefont {D.~V.}\ \bibnamefont
  {Schroeder}},\ }\href@noop {} {\emph {\bibinfo {title} {An Introduction to
  Quantum Field Theory}}}\ (\bibinfo  {publisher} {Westview Press},\ \bibinfo
  {address} {Boulder},\ \bibinfo {year} {1995})\BibitemShut {NoStop}%
\bibitem [{\citenamefont {Eberlein}\ and\ \citenamefont
  {Robaschik}(2006)}]{PRDEberleinRobaschik}%
  \BibitemOpen
  \bibfield  {author} {\bibinfo {author} {\bibfnamefont {C.}~\bibnamefont
  {Eberlein}}\ and\ \bibinfo {author} {\bibfnamefont {D.}~\bibnamefont
  {Robaschik}},\ }\href {\doibase 10.1103/PhysRevD.73.025009} {\bibfield
  {journal} {\bibinfo  {journal} {Phys. Rev. D}\ }\textbf {\bibinfo {volume}
  {73}},\ \bibinfo {pages} {025009} (\bibinfo {year} {2006})}\BibitemShut
  {NoStop}%
\bibitem [{\citenamefont {Johnson}\ and\ \citenamefont
  {Lippmann}(1949)}]{JohnsonLippmann}%
  \BibitemOpen
  \bibfield  {author} {\bibinfo {author} {\bibfnamefont {M.~H.}\ \bibnamefont
  {Johnson}}\ and\ \bibinfo {author} {\bibfnamefont {B.~A.}\ \bibnamefont
  {Lippmann}},\ }\href {\doibase 10.1103/PhysRev.76.828} {\bibfield  {journal}
  {\bibinfo  {journal} {Phys. Rev.}\ }\textbf {\bibinfo {volume} {76}},\
  \bibinfo {pages} {828} (\bibinfo {year} {1949})}\BibitemShut {NoStop}%
\bibitem [{\citenamefont {Carniglia}\ and\ \citenamefont
  {Mandel}(1971)}]{CarnigliaMandel}%
  \BibitemOpen
  \bibfield  {author} {\bibinfo {author} {\bibfnamefont {C.~K.}\ \bibnamefont
  {Carniglia}}\ and\ \bibinfo {author} {\bibfnamefont {L.}~\bibnamefont
  {Mandel}},\ }\href {\doibase 10.1103/PhysRevD.3.280} {\bibfield  {journal}
  {\bibinfo  {journal} {Phys. Rev. D}\ }\textbf {\bibinfo {volume} {3}},\
  \bibinfo {pages} {280} (\bibinfo {year} {1971})}\BibitemShut {NoStop}%
\bibitem [{\citenamefont {Hammer}(2003)}]{Hammer}%
  \BibitemOpen
  \bibfield  {author} {\bibinfo {author} {\bibfnamefont {H.}~\bibnamefont
  {Hammer}},\ }\href
  {http://www.tandfonline.com/doi/abs/10.1080/09500340308235171#.Ud61Ez54btO}
  {\bibfield  {journal} {\bibinfo  {journal} {J. Mod. Opt.}\ }\textbf {\bibinfo
  {volume} {50}},\ \bibinfo {pages} {207} (\bibinfo {year} {2003})}\BibitemShut
  {NoStop}%
\bibitem [{\citenamefont {Glauber}\ and\ \citenamefont
  {Lewenstein}(1991)}]{GlauberLewenstein}%
  \BibitemOpen
  \bibfield  {author} {\bibinfo {author} {\bibfnamefont {R.~J.}\ \bibnamefont
  {Glauber}}\ and\ \bibinfo {author} {\bibfnamefont {M.}~\bibnamefont
  {Lewenstein}},\ }\href {\doibase 10.1103/PhysRevA.43.467} {\bibfield
  {journal} {\bibinfo  {journal} {Phys. Rev. A}\ }\textbf {\bibinfo {volume}
  {43}},\ \bibinfo {pages} {467} (\bibinfo {year} {1991})}\BibitemShut
  {NoStop}%
\bibitem [{\citenamefont {Elson}\ and\ \citenamefont
  {Ritchie}(1971)}]{ElsonRitchie}%
  \BibitemOpen
  \bibfield  {author} {\bibinfo {author} {\bibfnamefont {J.~M.}\ \bibnamefont
  {Elson}}\ and\ \bibinfo {author} {\bibfnamefont {R.~H.}\ \bibnamefont
  {Ritchie}},\ }\href {\doibase 10.1103/PhysRevB.4.4129} {\bibfield  {journal}
  {\bibinfo  {journal} {Phys. Rev. B}\ }\textbf {\bibinfo {volume} {4}},\
  \bibinfo {pages} {4129} (\bibinfo {year} {1971})}\BibitemShut {NoStop}%
\bibitem [{\citenamefont {Barton}(1997)}]{BartonPlasma}%
  \BibitemOpen
  \bibfield  {author} {\bibinfo {author} {\bibfnamefont {G.}~\bibnamefont
  {Barton}},\ }\href {\doibase 10.1098/rspa.1997.0132} {\bibfield  {journal}
  {\bibinfo  {journal} {Proc. R. Soc. Lond. A}\ }\textbf {\bibinfo {volume}
  {453}},\ \bibinfo {pages} {2461} (\bibinfo {year} {1997})}\BibitemShut
  {NoStop}%
\bibitem [{\citenamefont {Eberlein}\ and\ \citenamefont
  {Zietal}(2012{\natexlab{a}})}]{RJZarxiv}%
  \BibitemOpen
  \bibfield  {author} {\bibinfo {author} {\bibfnamefont {C.}~\bibnamefont
  {Eberlein}}\ and\ \bibinfo {author} {\bibfnamefont {R.}~\bibnamefont
  {Zietal}},\ }\href {\doibase 10.1103/PhysRevA.86.022111} {\bibfield
  {journal} {\bibinfo  {journal} {Phys. Rev. A}\ }\textbf {\bibinfo {volume}
  {86}},\ \bibinfo {pages} {022111} (\bibinfo {year}
  {2012}{\natexlab{a}})}\BibitemShut {NoStop}%
\bibitem [{\citenamefont {Bennett}(2013)}]{REBthesis}%
  \BibitemOpen
  \bibfield  {author} {\bibinfo {author} {\bibfnamefont {R.}~\bibnamefont
  {Bennett}},\ }\href@noop {} {\bibfield  {journal} {\bibinfo  {journal}
  {University of Sussex Doctoral Thesis}\ } (\bibinfo {year}
  {2013})}\BibitemShut {NoStop}%
\bibitem [{\citenamefont {Lifshitz}\ and\ \citenamefont
  {Pitaevskii}(1980)}]{Lifshitz}%
  \BibitemOpen
  \bibfield  {author} {\bibinfo {author} {\bibfnamefont {E.~M.}\ \bibnamefont
  {Lifshitz}}\ and\ \bibinfo {author} {\bibfnamefont {L.~P.}\ \bibnamefont
  {Pitaevskii}},\ }\href@noop {} {\emph {\bibinfo {title} {Statistical Physics,
  Part 2}}}\ (\bibinfo  {publisher} {Pergamon},\ \bibinfo {address} {Oxford},\
  \bibinfo {year} {1980})\ \bibinfo {note} {chapter VIII}\BibitemShut {NoStop}%
\bibitem [{\citenamefont {Maradudin}\ and\ \citenamefont
  {Mills}(1975)}]{MaradudinMills}%
  \BibitemOpen
  \bibfield  {author} {\bibinfo {author} {\bibfnamefont {A.~A.}\ \bibnamefont
  {Maradudin}}\ and\ \bibinfo {author} {\bibfnamefont {D.~L.}\ \bibnamefont
  {Mills}},\ }\href {\doibase 10.1103/PhysRevB.11.1392} {\bibfield  {journal}
  {\bibinfo  {journal} {Phys. Rev. B}\ }\textbf {\bibinfo {volume} {11}},\
  \bibinfo {pages} {1392} (\bibinfo {year} {1975})}\BibitemShut {NoStop}%
\bibitem [{\citenamefont {Eberlein}\ and\ \citenamefont
  {Zietal}(2012{\natexlab{b}})}]{RJZanisotropic}%
  \BibitemOpen
  \bibfield  {author} {\bibinfo {author} {\bibfnamefont {C.}~\bibnamefont
  {Eberlein}}\ and\ \bibinfo {author} {\bibfnamefont {R.}~\bibnamefont
  {Zietal}},\ }\href {\doibase 10.1103/PhysRevA.86.062507} {\bibfield
  {journal} {\bibinfo  {journal} {Phys. Rev. A}\ }\textbf {\bibinfo {volume}
  {86}},\ \bibinfo {pages} {062507} (\bibinfo {year}
  {2012}{\natexlab{b}})}\BibitemShut {NoStop}%
\bibitem [{\citenamefont {Babiker}(1976)}]{BabikerPlasma}%
  \BibitemOpen
  \bibfield  {author} {\bibinfo {author} {\bibfnamefont {M.}~\bibnamefont
  {Babiker}},\ }\href {\doibase 10.1103/PhysRevB.13.3056} {\bibfield  {journal}
  {\bibinfo  {journal} {Phys. Rev. B}\ }\textbf {\bibinfo {volume} {13}},\
  \bibinfo {pages} {3056} (\bibinfo {year} {1976})}\BibitemShut {NoStop}%
\bibitem [{\citenamefont {Walter}\ and\ \citenamefont
  {Cohen}(1972)}]{SiliconDielectricFunction}%
  \BibitemOpen
  \bibfield  {author} {\bibinfo {author} {\bibfnamefont {J.~P.}\ \bibnamefont
  {Walter}}\ and\ \bibinfo {author} {\bibfnamefont {M.~L.}\ \bibnamefont
  {Cohen}},\ }\href {\doibase 10.1103/PhysRevB.5.3101} {\bibfield  {journal}
  {\bibinfo  {journal} {Phys. Rev. B}\ }\textbf {\bibinfo {volume} {5}},\
  \bibinfo {pages} {3101} (\bibinfo {year} {1972})}\BibitemShut {NoStop}%
\bibitem [{\citenamefont {{G{\'o}mez Rivas}}\ \emph {et~al.}(2006)\citenamefont
  {{G{\'o}mez Rivas}}, \citenamefont {{Kuttge}}, \citenamefont {{Kurz}},
  \citenamefont {{Haring Bolivar}},\ and\ \citenamefont
  {{S{\'a}nchez-Gil}}}]{SemiconductorPlasma}%
  \BibitemOpen
  \bibfield  {author} {\bibinfo {author} {\bibfnamefont {J.}~\bibnamefont
  {{G{\'o}mez Rivas}}}, \bibinfo {author} {\bibfnamefont {M.}~\bibnamefont
  {{Kuttge}}}, \bibinfo {author} {\bibfnamefont {H.}~\bibnamefont {{Kurz}}},
  \bibinfo {author} {\bibfnamefont {P.}~\bibnamefont {{Haring Bolivar}}}, \
  and\ \bibinfo {author} {\bibfnamefont {J.~A.}\ \bibnamefont
  {{S{\'a}nchez-Gil}}},\ }\href {\doibase 10.1063/1.2177348} {\bibfield
  {journal} {\bibinfo  {journal} {Appl. Phys. Lett.}\ }\textbf {\bibinfo
  {volume} {88}},\ \bibinfo {pages} {082106} (\bibinfo {year}
  {2006})}\BibitemShut {NoStop}%
\bibitem [{\citenamefont {Cohen-Tannoudji}\ \emph {et~al.}(1977)\citenamefont
  {Cohen-Tannoudji}, \citenamefont {Diu},\ and\ \citenamefont
  {Lalo{\"e}}}]{CohenTannoudji}%
  \BibitemOpen
  \bibfield  {author} {\bibinfo {author} {\bibfnamefont {C.}~\bibnamefont
  {Cohen-Tannoudji}}, \bibinfo {author} {\bibfnamefont {B.}~\bibnamefont
  {Diu}}, \ and\ \bibinfo {author} {\bibfnamefont {F.}~\bibnamefont
  {Lalo{\"e}}},\ }\href@noop {} {\emph {\bibinfo {title} {Quantum Mechanics}}}\
  (\bibinfo  {publisher} {Wiley},\ \bibinfo {address} {New York},\ \bibinfo
  {year} {1977})\ \bibinfo {note} {section ${\rm E}_{\rm VI}$}\BibitemShut
  {NoStop}%
\end{thebibliography}

%

\end{document}